\documentclass{article}
\usepackage{enumitem}
\usepackage{microtype}
\usepackage{graphicx}
\usepackage{subfigure}
\usepackage{booktabs} 
\usepackage{makecell}
\usepackage{hyperref}
\usepackage{multirow}
\usepackage{tikz}
\usetikzlibrary{shapes.geometric, calc, backgrounds, positioning, shadows, fit}

\usepackage{xcolor} 
\usepackage{xspace}

\usepackage{algorithmic}
\definecolor{OliveGreen}{RGB}{107,142,35} 
\renewcommand{\algorithmiccomment}[1]{\hfill {\color{OliveGreen!75} \(\triangleright\) #1}}
\usepackage[toc]{appendix} 
\usepackage[accepted]{icml2025}

\usepackage{amsmath}
\usepackage{amssymb}
\usepackage{mathtools}
\usepackage{amsthm}
\usepackage{rotating} 
\usepackage{booktabs} 
\usepackage{pifont}

\usepackage[capitalize,noabbrev]{cleveref}

\definecolor{olmodna}{RGB}{128,64,128}
\definecolor{groundtruth}{RGB}{53,122,235}
\definecolor{nttransformer}{RGB}{72,155,125}
\definecolor{nttransformer2}{RGB}{127,127,127}
\definecolor{darkgreen}{RGB}{0, 100, 0} 
\definecolor{lightgrey}{RGB}{211, 211, 211}
\definecolor{lightred}{RGB}{255, 182, 193}
\definecolor{lightgreenRGB}{RGB}{144, 238, 144}
\theoremstyle{plain}

\theoremstyle{definition}

\theoremstyle{remark}

\newcommand{\method}{Omni-DNA\xspace}

\usepackage[textsize=tiny]{todonotes}

\icmltitlerunning{Omni-DNA: A Unified Genomic Foundation Model for Cross-Modal and Multi-Task Learning}

\begin{document}

\twocolumn[

\icmltitle{Omni-DNA: A Unified Genomic Foundation Model for Cross-Modal and Multi-Task Learning}


\icmlsetsymbol{equal}{*}

\begin{icmlauthorlist}
\icmlauthor{Zehui Li}{zzz,yyy}
\icmlauthor{Vallijah Subasri}{vec,UHN}
\icmlauthor{Yifei Shen}{zzz}
\icmlauthor{Dongsheng Li}{zzz}
\icmlauthor{Yiren Zhao}{yyy}
\icmlauthor{Guy-Bart Stan}{yyy}
\icmlauthor{Caihua Shan}{zzz}
\end{icmlauthorlist}

\icmlaffiliation{yyy}{Imperial College London}
\icmlaffiliation{zzz}{Microsoft Research}
\icmlaffiliation{vec}{Vector Institute}
\icmlaffiliation{UHN}{University Health Network}

\icmlcorrespondingauthor{Zehui Li}{zl6222@ic.ac.uk}
\icmlcorrespondingauthor{Caihua Shan}{caihua.shan@microsoft.com}


\vskip 0.3in
]



\printAffiliationsAndNotice{} 

\begin{abstract}

Large Language Models (LLMs) demonstrate remarkable generalizability across diverse tasks, yet genomic foundation models (GFMs) still require separate finetuning for each downstream application, creating significant overhead as model sizes grow.
Moreover, existing GFMs are constrained by rigid output formats, limiting their applicability to various genomic tasks. In this work, we revisit the transformer-based auto-regressive models and introduce \textsf{\method}, a family of cross-modal multi-task models ranging from 20 million to 1 billion parameters.  Our approach consists of two stages: (i) \textbf{pretraining} on DNA sequences with next token prediction objective, and (ii) \textbf{expanding} the multi-modal task-specific tokens and \textbf{finetuning} for multiple downstream tasks simultaneously. When evaluated on the Nucleotide Transformer and GB benchmarks, \method achieves state-of-the-art performance on 18 out of 26 tasks. Through multi-task finetuning, \method addresses 10 acetylation and methylation tasks at once, surpassing models trained on each task individually. Finally, we design two complex genomic tasks, \textit{DNA2Function} and \textit{Needle-in-DNA}, which map DNA sequences to textual functional descriptions and images, respectively, indicating Omni-DNA’s cross-modal capabilities to broaden the scope of genomic applications. All the models are available through \url{https://huggingface.co/collections/zehui127}.

\end{abstract}

\section{Introduction}
\begin{figure}

\centering
\resizebox{\columnwidth}{!}{
\begin{tikzpicture}
  \tikzstyle{every node}=[font=\small]
  \sffamily

  \pgfdeclarelayer{-1}
  \pgfsetlayers{-1,main}
  \tikzset{zlevel/.style={%
        execute at begin scope={\pgfonlayer{#1}},
        execute at end scope={\endpgfonlayer}
      }}

  \begin{scope}[zlevel=main]
    \node[draw=none, inner sep=0, text width=\linewidth, align=left] at (0, 0) (input1) {
      \textbf{Model Input:} What is the function of this sequence?GGCTG...TTTTCTGA 
    };

    \node[draw=none, inner sep=0, text width=\linewidth, align=left, below=0.2cm of input1.south, anchor=north] (truth1) {
      \textbf{Ground Truth:} \colorbox{lightgrey}{Olfactory receptor} 5A2-like (OR5A2L) is a member of the olfactory receptor gene family, which is involved in the \colorbox{lightgrey}{detection of volatile substances}, contributing to the sense of smell.
    };

    \node[draw=none, inner sep=0, text width=\linewidth, align=left, below=0.2cm of truth1.south, anchor=north] (olmo1) {
      {\color{olmodna}\textbf{\method}}: The mRNA for \colorbox{lightgreenRGB}{olfactory receptor family 5} subfamily \colorbox{lightred}{B member 109 ( OR5B109 )} encodes a protein that is part of the olfactory receptor family. These receptors are involved in the \colorbox{lightgreenRGB}{detection of odorant molecules} and and play a crucial role in the sense of smell.
    };





  \end{scope}

  \begin{scope}[zlevel=-1]
    \node[draw=black, fill=white, inner sep=0.25cm, rounded corners, drop shadow, fit={(input1) (truth1) (olmo1)}] (frame) {};
  \end{scope}

  \begin{scope}[zlevel=-1]
    \node[draw=none, fill=none, inner sep=0.075cm, fit={(frame)}] {};
  \end{scope}
\end{tikzpicture}
}
\vspace{-2em} 
\caption{\textbf{Demonstration of {\color{olmodna}\textsf{\method}}'s cross-modal capabilities.} Given a DNA sequence, \method could generate a natural language description for functional annotations.}
\label{fig:model-output}
\vspace{-2em} 
\end{figure}


The volume of genomic data has been increasing at an exponential rate over the past decades~\cite{lathe2008genomic}, making it infeasible to annotate these sequences manually. 
Genomic Foundation Models (GFMs)~\cite{nguyen2024hyenadna,zhou2023dnabert,dalla2024nucleotide,schiff2024caduceus}, a type of Deep Neural Networks (DNNs) for genomic sequence modeling, has emerged as essential tools to automate the annotation process. GFMs has been used for associating mutation in the genome with diseases~\cite{benegas2023dna,cheng2023accurate}, revealing regulatory effects of genomic sequences~\cite{avsec2021effective,linder2025predicting}, and genomic elements annotation~\cite{de2024segmentnt}. Although these GFMs achieve great performance on genomic tasks~\citep{zhou2015predicting,grevsova2023genomic}, they are still far away from generalist models, which can handle multiple tasks simultaneously. This contrasts with Large Language Models (LLMs)~\citep{radford2019language,team2023gemini,touvron2023llama}, which have seen remarkable success at solving general tasks from question answering to theorem proofs~\cite{xin2024deepseek}. The success of LLMs is pretraining transformer-based auto-regressive models~\cite{vaswani2017attention} on internet-scale data, followed by post-training to increase instruction-following abilities~\cite{lambert2024t}. Inspired by this, we ask: \textit{is it possible to develop a generalist genomic foundation model for addressing diverse genomic tasks simultaneously?}

Existing GFMs vary significantly in architecture and tokenization strategies, but they all follow a ``pretrain + task-specific finetune'' paradigm: first pretraining on unlabeled DNA sequences, and task-specific Multi-Layer Perceptrons (MLPs) are attached for finetuning. This paradigm limits the generalization of the model in two ways. \textbf{(i)} The pretrained model needs to be separately finetuned $K$ times given $K$ tasks. This results in the additional cost of storing $ \mathcal{O}(K) $ copies of model weights, incurring significant I/O latency, memory costs, and context-switching penalties. This also prevents models from leveraging shared biological patterns (e.g., conserved regulatory motifs or chromatin accessibility signatures~\cite{van2012motif, yang2023pioneer}) across finetuning tasks. \textbf{(ii)} The output space of existing models is rigidly constrained to DNA sequences or predefined genomic labels, limiting their ability to generate diverse, semantically meaningful outputs. As a result, current GFMs struggle for multi-modal downstream tasks such as DNA2Text and DNA2Image.


To overcome these limitations, we introduce \textsf{\method}, a family of cross-modal multi-task models ranging from 20 million to 1 billion parameters. This is achieved through carefully ablated pretraining of transformer-based auto-regressive models on DNA sequences and cross-modal multi-task finetuning. 

The \textit{pretraining} stage explores previously overlooked key configurations and their impact on training dynamics. This includes a comparison of non-parametric LayerNorm~\cite{ba2016layer} vs. RMSNorm~\cite{zhang2019root}, RoPE~\cite{su2024roformer} vs. ALiBi~\cite{press2021train} positional embeddings, and other configurations, ultimately leading \method to achieve state-of-the-art performance on 18 out of 26 tasks on the Nucleotide Transformer~\cite{dalla2024nucleotide} and GB benchmarks~\cite{grevsova2023genomic}. When evaluated with a conventional paradigm, \method outperforms existing bidirectional models.

\begin{figure}[ht!]
    \centering
    \includegraphics[width=0.7 \columnwidth]{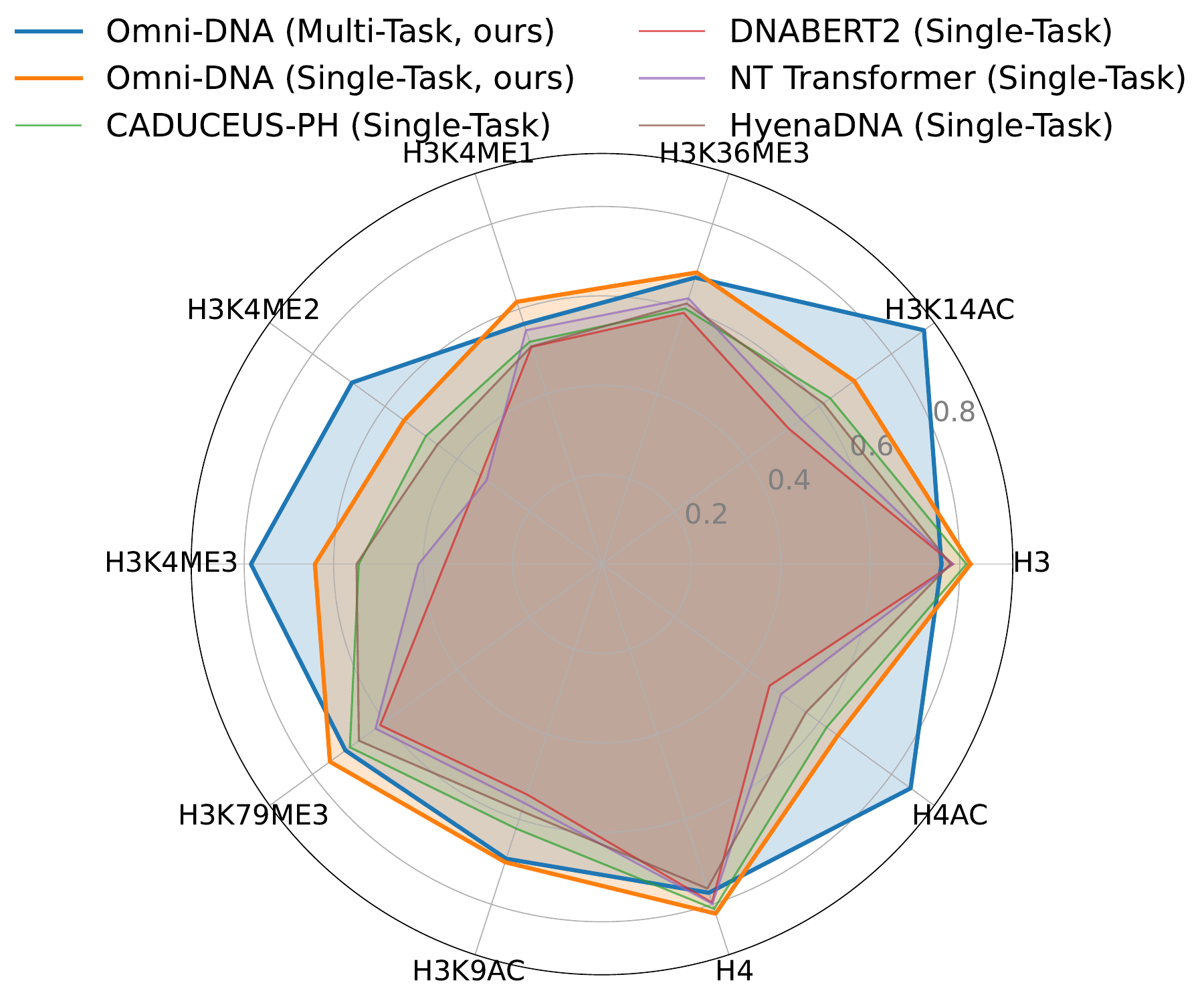}
     \vspace{-1em} 
    \caption{\textbf{Accuracy comparison of Omni-DNA@mult. against Omni-DNA@sgl. and baselines across 10 NT tasks.} Omni-DNA@mult. achieves the highest average accuracy.}
    \label{fig:multi-tasking}
    \vspace{-1em}
\end{figure}

\textit{Cross-modal multi-task finetuning} expands the vocabulary in the model's tokenizer by dynamically modifying the embedding matrix. To address the distribution shift of existing tokens due to vocabulary expansion~\cite{hewitt2021initializing}, we propose \textit{Important Key Repetition} and adopt NEFTune~\cite{jain2023neftune}. Additionally, we implement a task unification strategy, aligning multiple genomic tasks under a unified format. The combination of these techniques enables a single model to solve multiple acetylation and methylation tasks at once, surpassing models trained on each task individually, as demonstrated in \cref{fig:multi-tasking}. Furthermore, beyond conventional genomic tasks, \method explores direct mapping from DNA sequences to functional descriptions and images, unlocking cross-modal capabilities.

\textbf{In summary, our contributions are three-fold}: \\
\noindent $(1)$  We revisit the auto-regressive transformers for GFMs, indicating the potential of next token prediction paradigm.  \\
\noindent$(2)$  We present \method, a unified cross-modal and multi-task GFM ranging from 20M to 1B parameters, achieving SOTA performance on multiple conventional benchmarks.\\ 
\noindent$(3)$ We evaluate \method on novel cross-modal and multi-task genomic tasks, such as DNA2Func and DNA2Image.

\section{Preliminaries and Notations}


\subsection{Genomic Sequence Modeling} 
DNA is a polymer made of up four types of nucleotides: Adenine (\textit{A}), Thymine (\textit{T}), Guanine (\textit{G}), and Cytosine (\textit{C}). Let $ \mathbb{N}_4 = \{A, T, G, C\}$. A DNA sequence of length $T$, denoted as $\mathbf{x} = {(\mathbf{x}_1, \mathbf{x}_2,...,\mathbf{x}_T)} \in \mathbb{N}^T_4$, follows a natural distribution $\mathbf{x} \sim p_\theta(\mathbf{x})$. We use $p_{\hat{\theta}}(\mathbf{x})$ to represent an estimate to the true distribution. The dataset of unlabeled genomic sequences is given in the form of \( \{\mathbf{x}^{(i)}\}_{i=1}^N\).

Genomic sequence modeling aims to learn a function $f$ that maps input sequences to biological annotations using a labeled dataset \(\mathcal{D} = \{\mathbf{x}^{(i)}, \mathbf{y}^{(i)}\}_{i=1}^N\). The type of $y^{(i)}$ varies depending on the types of tasks: $y^{(i)}$ is a class label in \textbf{DNA Sequence Classification}~\cite{grevsova2023genomic,dalla2024nucleotide}. Or a real value vector in \textbf{Genomic Assay Prediction} tasks~\cite{avsec2021effective,linder2025predicting}. Current genomic sequence models typically follow a two-stage strategy. In the pretraining phase, we learn the data distribution $p_{\hat{\theta}}(x)$ on a unlabeled dataset from unlabeled data using losses such as masked language modeling (MLM)
$p(\mathbf{x}) = \prod_{t \in \mathcal{M}} p_\theta(\mathbf{x}_t \mid \mathbf{x}_{1:t-1}, \mathbf{x}_{t+1:T})$ 
or next token prediction (NTP) $p(\mathbf{x}) = \prod_{t=1}^{T}p_\theta(\mathbf{x}_t | \mathbf{x}_{1:t-1})$.

\subsection{Supervised Finetuning}

Supervised Finetuning (SFT) plays a key role in enhancing the instruction-following~\citep{mishra2021reframing,sanh2021multitask,wei2022chain} and reasoning capabilities~\citep{lambert2024t}. For a pretrained auto-regressive model with a fixed vocabulary \( V_x \) and a labeled dataset  $\mathcal{D}$, SFT maximizes the likelihood:
$
\hat{\theta} = \arg\max_{\theta} \sum_{i=1}^N \log p_\theta\left(\mathbf{y}^{(i)} \mid \mathbf{x}^{(i)}\right).
$
This process retains the model’s pretrained knowledge while aligning its outputs with task-specific objectives, typically using \textbf{cross-entropy loss} on the target tokens:
\begin{equation} \label{eq:3} \small
    p_\theta(\mathbf{y}^{(i)} | \mathbf{x}^{(i)}) = \sum_{i=1}^N \sum_{t=1}^{T'} \log p_\theta(\mathbf{y}_t^{(i)} | \mathbf{y}^{(i)}_{1:t-1},\mathbf{x}^{(i)}).
\end{equation}
Notably, $\mathbf{y}$ could include new set of vocabulary $V_{z}$ that do not overlap with $V_{o}$. Therefore, each term in \cref{eq:3} is computed through \Cref{eq:4}.
\begin{equation} \label{eq:4} 
\small
    p_\theta(\mathbf{y}_t \mid \mathbf{y}_{1:t-1},\mathbf{x}) = \frac{\exp(h_{t-1}^\top e_{\mathbf{y}_t})}{\sum\limits_{m \in V_{o}} \exp(h_{t-1}^\top e_m) + \sum\limits_{n \in V_{z}} \exp(h_{t-1}^\top e_n)},
\end{equation} 
where $h_{t-1}$ is the neural representation of the prefix sequence $(\mathbf{y}_{1:t-1},\mathbf{x})$, $e_m$ is the embedding of vocab $m$.

As a result, during the finetuning stage, the embeddings for each new vocabulary token \( n \in V_z \) must be initialized, and the original token probabilities are shifted due to the expanded output space, as detailed in~\citet{hewitt2021initializing}.



\section{Approach}
\begin{figure}[t!]
    \centering
    \includegraphics[width=\columnwidth]{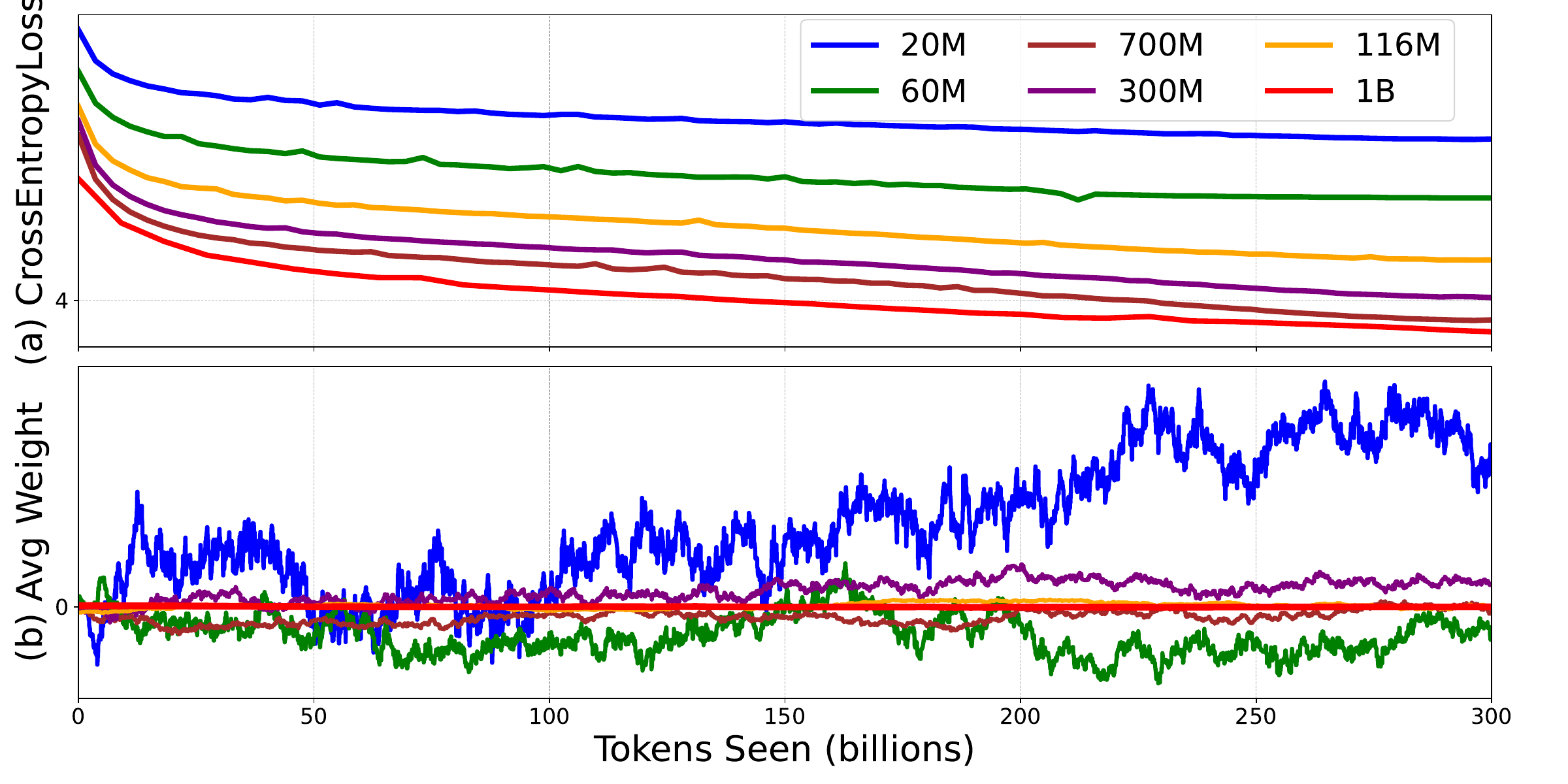}
    \vspace{-2em} 
    \caption{\textbf{(a) Cross Entropy Loss} on test set during pretraining. The models with varying sizes show a stable decrease in loss. \\
    \textbf{(b) No-Bias Normalization Layer} stabilizes the average value of feed-forward weights in transformer layers. This pattern is consistent across all the transformer blocks.}
     \vspace{-1em}
    \label{fig:no-bais-norm}
\end{figure}
 \begin{figure*}[ht!]
    \centering
    \includegraphics[width=\linewidth]{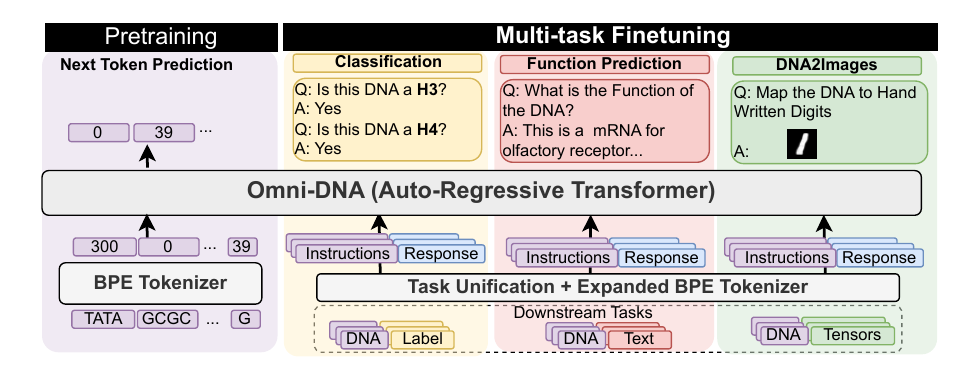}
    \vspace{-2em} 
    \caption{\textbf{Overview of of Omni-DNA architecture.} In  \textbf{pretraining}, \method are trained on DNA only with next-token prediction. \textbf{Multi-task finetuning} enables the model to perform diverse tasks including classification, function prediction, and DNA-to-image.}
    \label{fig:overview_method}
  \vspace{-1em}
\end{figure*}

In this section, we introduce \method with various model sizes for genomic tasks. The pretraining process is present in Section~\ref{sec:pretraining}, followed by cross-modal multi-task finetuning in Section~\ref{sec:omni-finetuning}. The framework is illustrated in Figure~\ref{fig:overview_method}.


\subsection{Pretraining}
 
\label{sec:pretraining}
\paragraph{Pretraining Dynamics:} we pretrain a series of auto-regressive transformer-based models using unlabeled DNA data, with log-uniformly spaced sizes ranging from 20 Million to 1 Billion. The pretraining procedure are conducted with minimum loss spike as shown in \cref{fig:no-bais-norm}(a) across models with various sizes. These model are trained on 300 billion nucleotides. The full hyperparameters used by our model and a comparison with prior genomic models are shown in \cref{app:pretrain-config}, below highlights key configurations.

 
\paragraph{Model Architecture:} We build on Open Language Model (OLMo)~\citep{groeneveld2024olmo} for DNA Sequence Modeling. Compared to the vanilla transformer~\citep{vaswani2017attention},  several improvements introduced by LLaMA family~\citep{touvron2023llama} and OLMo family are adopted to our model. 
    \textbf{(i) LayerNorm} The Non-parametric layer norm~\citep{ba2016layer} without Bias are applied to our 116M and 1B model. RMSNorm~\citep{zhang2019root} are applied to the remaining fours models. We find while both types of LayerNorm result in a stable pretraining process, the Non-parametric layer norm tends to maintain more stable average weights on the feed forward layer as shown in \cref{fig:no-bais-norm}, which in turns help to achieve higher accuracy when performing fullsize finetuning on the downstream tasks (\cref{sec:fullfinetune}).
    \textbf{(ii) Relative Positional Embeddings} 
    Although previous work~\cite{zhou2023dnabert} indicates that relative positional embedding methods such as ALiBi~\cite{press2021train} may offer improved extrapolation capabilities, the context length of the pretrained model can still be easily extended during finetuning. We observed slower convergence when using ALiBi in our pretraining experiments (see \cref{sec:ablation}).
    \textbf{(iii) BPE Tokenizer} We adopt Byte-Pair Encoding (BPE)~\cite{sennrich2015neural} with an initial vocabulary of 4096, following~\citet{zhou2023dnabert}. During the vocabulary expansion, newly added tokens are always recognized as standalone tokens rather than retraining the tokenizer.
    

\paragraph{Pretraining Dataset Deduplication}

Duplicate data can slow model convergence, making deduplication standard in LLM pretraining~\cite{rae2021scaling, groeneveld2024olmo}. Because genomic data is highly duplicated, we removed exact duplicates from NCBI's  multi-species genome dataset~\cite{schoch2020ncbi}, reducing it to 30 billion nucleotides. Despite this reduction, multiple epochs can still be used in training.


\begin{table*}[t!]
\centering
\footnotesize 
\caption{Examples of Genomic Task Unification, including inputs, labels, instructions, and responses.}
\resizebox{\textwidth}{!}{
\begin{tabular}{@{}lcc|cc@{}}
\toprule
\textbf{Task Type} & \multicolumn{2}{c|}{\textbf{Original}} & \multicolumn{2}{c}{\textbf{Unified Format}} \\ 
\cmidrule(lr){2-3} \cmidrule(lr){4-5}
 & \textbf{Input} $x^{(i)}$& \textbf{Target} $y^{(i)}$ & \textbf{Instruction} $\tilde{x}^{(i)}$ & \textbf{Response} $\tilde{y}^{(i)}$ \\ \midrule
\small Promoter CLS\fontsize{5}{5}\selectfont~\yrcite{dalla2024nucleotide}& AT...T & 0/1 & AT...T\textcolor{cyan!50!blue}{is this a promoter?} & Yes/No \\
\small  Enhancer Types CLS \fontsize{5}{5}\selectfont~\yrcite{dalla2024nucleotide}&  AT...A  &0/1/2  & AT...A\textcolor{cyan!50!blue}{what is enhancer type?} & Type 0/1/2 \\
\small  DNA2Image & AT...G & \includegraphics[width=0.4cm]{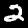} & AT...G\textcolor{cyan!50!blue}{map to image} & $[0,2,...,512]$ (discretized tokens) \\
\small  DNA2Text & AT...T & \small {It encodes H-box helicase protein} & AT...T\textcolor{cyan!50!blue}{what is the function?} & It encodes H-box helicase protein \\
\bottomrule
\end{tabular}
}
\label{tab:unification_tasks}
 \vspace{-1em}
\end{table*}

\subsection{Cross-modal Multi-task Finetuning}
\label{sec:omni-finetuning}

\begin{algorithm}[t!] \small
\caption{\textbf{Cross-modal Multi-task Finetuning}}
\begin{algorithmic}[1]
\REQUIRE $K$ datasets $\{\mathcal{D}_k\}_{k=1}^K$, 
         pretrained model $f_\theta$, 
         vocabulary $V_\text{pretrained}$, repeating factor $\alpha$, Noise Level $\beta$

\STATE $\mathcal{D}_\text{uni} \gets \bigcup_{k=1}^K \mathcal{D}_k$
\COMMENT{Task Unification}
\STATE Extract task-specific multi-modal tokens $V_k$ from $\mathcal{D}_k$
\STATE $V_\text{expand} \gets V_\text{pretrained} \cup \bigcup_{k=1}^K V_k$
\COMMENT{Vocab. expansion}

\REPEAT
    \STATE $\{(x^{(i)}, y^{(i)})\}_{i=1}^B \sim \mathcal{D}_\text{uni}$ \COMMENT{Batch size $B$}
    
    \FOR{$i$ from 1 to $B$}
        \STATE $y^{(i)} \gets \text{ReplicateKeyToken}\bigl(y^{(i)},\alpha \bigr)$ 
            \COMMENT{Replicate Class Labels in target for Classification Tasks}
        \STATE $\text{loss}^{(i)} \gets -\sum_{t=1}^{|y^{(i)}|} 
            \log p_\theta\Bigl(y^{(i)}_t \,\Big\vert\, y^{(i)}_{1:t-1},x^{(i)} ,\beta\Bigr)$
            \COMMENT{Add noise using NEFTune with $\beta$ during loss computing}
    \ENDFOR
    \STATE $\text{Loss}_\text{batch} \gets \frac{1}{B}\sum_{i=1}^{B} \text{loss}^{(i)}$
    \STATE $\theta \gets \text{optimizer}\bigl(\theta, \text{Loss}_\text{batch}\bigr)$ 
\UNTIL{\emph{stopping criterion} or \emph{max iterations}}

\label{algorithm:1}
\end{algorithmic}
\end{algorithm}

Here we extend pretrained models to handle multiple modalities beyond DNA tokens and enable learning of diverse downstream tasks through a single finetuning process. We first introduce the whole process of finetuning and then describe the detailed modules.

\paragraph{Whole Process} As illustrated in \cref{algorithm:1}, we need to finetune $K$ task-specific datasets with different token sets and modalities. We first merge $K$ task-specific datasets into a unified dataset $\mathcal{D}_\text{uni}$, and the tokenizer vocabulary are expanded to include unique new tokens from each task. During each training iteration, a batch of $(x^{(i)}, y^{(i)})\}_{i=1}^B$ is first sampled from $\mathcal{D}_\text{uni}$, $y^{(i)}$ is replicated by a factor $\alpha$ to emphasize class labels, and the loss is computed. The parameters $\theta$ are then updated via gradient-based optimization. By unifying multiple tasks under a single loop, the algorithm encourages the model improves the average performance across diverse genomic tasks. 

\paragraph{Multi-task Dataset Unification}  
We integrate \( K \) genomic tasks with labeled datasets \(\{\mathcal{D}_k\}_{k=1}^K\) of various formats into a unified dataset $\mathcal{D}_{\text{uni}} = \{\tilde{x}^{(i)},\tilde{y}^{(i)}\}$. Specifically, each sample from task $k$ is modified by appending a task-specific prompt \(\textcolor{cyan!50!blue}{\text{Prompt}_k}\) to its input \(x^{(i)}\):
$
\tilde{x}^{(i)} = \big(x^{(i)}, \textcolor{cyan!50!blue}{\text{Prompt}_k}\big).
$
The corresponding response \(\tilde{y}^{(i)}\) is treated as tokens and the typical transformation is shown in \cref{tab:unification_tasks}.


\paragraph{Vocabulary Expansion} As the multi-task dataset unification introduces new tokens (i.e., prompt and response tokens), it is necessary to expand the vocabulary during SFT stage. This is a key difference between genomic language model and normal language models. A direct result of vocabulary expansion is distribution shift. Based on~\cite{hewitt2021initializing}, the distribution shift after adding a new set of vocabulary  $V_{z}$ to pretrained model $\theta$ results in a new model $\theta'$, let $\scriptstyle Z = \sum\limits_{m \in V_{o}} \exp(h_{t-1}^\top e_m)$. Then the distribution shift are:
\begin{equation} \small
    p_{\theta'}(x_t \mid x_{1:t-1}) = p_\theta(x_t \mid x_{1:t-1}) \cdot \frac{1}{1 + \frac{\sum\limits_{n \in V_{z}} \exp(h_{t-1}^\top e_n)}{Z}}.
    \label{eq:5}
\end{equation}
 In other words, all the words probability are reduced. To alleviate it, we should minimize the size of added tokens. 

\paragraph{NEFTune \& Key Token Replication}  
To mitigate catastrophic forgetting~\cite{zheng2024towards} caused by distribution shifts from vocabulary expansion, we employ two simple yet effective techniques: \textit{NEFTune}~\citep{press2021train} and \textit{Key Token Replication}. NEFTune introduces random noise to embedding vectors during finetuning, reducing overfitting. Key Token Replication, applied to classification tasks, replicates label tokens in \( y^{(i)} \) by a factor of \( \alpha \). This ensures stronger signal propagation, facilitating the model’s transition from DNA sequence generation to classification by reinforcing the association between input sequences and their corresponding labels.

\begin{table*}[ht!]
\caption{\textbf{Comparison of Omni-DNA with existing GFMs on 18 NT downstream tasks.} The family of Omni-DNA obtains the best average result.
In splice tasks, Omni-DNA performs lower than NT models but still surpasses other models.}
\label{tab:pretraining_comparison}
\resizebox{\textwidth}{!}{
\begin{tabular}{@{}lllcccccc@{}}
\toprule
\tiny{\textbf{Pretraining}} & \tiny\textbf{Pretraining} & \textbf{Model} & \textbf{Histone} & \textbf{Enhancer} & \textbf{Promoter} & \textbf{Splice} & \textbf{Average} \\
\tiny\textbf{Objective} & \tiny\textbf{Data} & & \tiny (AVG. MCC in 10 Tasks)$\uparrow$ & \tiny (AVG. MCC in 2 Tasks)$\uparrow$ & \tiny (AVG. F1 in 3 Tasks)$\uparrow$ & \tiny (AVG. F1 in 3 Tasks)$\uparrow$ & \tiny (Across 18 Tasks)$\uparrow$ \\
\midrule
\multirow{5}{*}{\small MLM} & \multirow{2}{*}{\small\shortstack{Human}} 
                     & \small CADUCEUS-PH (1.9M)    & 0.635 & 0.4925 & 0.964 & 0.942 & 0.715 \\
                     &                                & \small CADUCEUS-PS (1.9M)   & 0.585  & 0.454 & 0.964 & 0.912 & 0.689 \\
\cmidrule{2-8}
                     & \multirow{6}{*}{\small\shortstack{Multi-\\Species}} 
                     & \small DNABERT2 (120M)       & 0.551 & 0.470 & 0.966 & 0.959 & 0.679 \\
                     &                                & \small NT (50M)      & 0.493 & 0.465 & 0.953 & 0.980 & 0.648 \\
                     &                                & \small NT (100M)  & 0.496 & 0.475 & 0.957 & 0.983 & 0.661 \\
                     &                                & \small NT (250M)          & 0.536 & 0.472 & 0.968 & 0.983 & 0.675 \\
                     &                                & \small NT (500M)          & 0.572 & 0.486 & 0.972 & \underline{0.983}   & 0.698 \\
                     &                                & \small NT (2.5B)     & 0.584 & 0.527 & \underline{0.971} & \textbf{0.986}  & 0.709 \\
\midrule
\multirow{8}{*}{\small NTP} & \multirow{1}{*}{\small\shortstack{Human}} 
                     & \small HyenaDNA (1.6M)     & 0.610 & 0.452 & 0.954 &0.954   & 0.707 \\
\cmidrule{2-8}
                     & \multirow{6}{*}{\small\shortstack{Multi-\\Species\\(\textbf{Ours})}} 
                     & \small Omni-DNA (20M)     & 0.538 & 0.484 & 0.957 & 0.897 & 0.662 \\
                     &                                & \small Omni-DNA (60M)      &0.559 & 0.499 & 0.966 & 0.938 & 0.690 \\
                     &                                & \small Omni-DNA (116M)  & \underline{0.675} & \underline{0.545} & 0.970 & 0.960   & \underline{0.755} \\
                     &                                & \small Omni-DNA (300M)    & 0.659 & 0.486 & 0.969    & 0.956& 0.741 \\
                     &                                & \small Omni-DNA (700M)       & 0.675 & 0.519 & 0.968  & 0.960 & 0.754 \\
                     &                                & \small Omni-DNA (1B)        & \textbf{0.694} & \textbf{0.536} & \textbf{0.973}  & 0.958  & \textbf{0.767} \\
\bottomrule
\end{tabular}
}
\vspace{-1em}
\end{table*}

\paragraph{Response Discretization} In many genomic tasks such as genomic assay prediction~\cite{avsec2021effective} and structure prediction~\cite{abramson2024accurate}, the response is a high-dimensional continuous variable $\mathbf{y} \in \mathbb{R}^{d_1 \times d_2\cdots \times d_k}$. To adopt Omni-DNA, we need to discretize the response first. 


Here we consider a three-dimensional continuous variable $\mathbf{y} \in R^{d_1\times d_2\times d_3}$. \textit{Response discretization} is achieved by training a VQ-VAE~\cite{van2017neural}, which learns a latent embedding space of dimension $[K, D]$, where $K$ represents the number of quantized embeddings, and $D$ denotes the dimensionality of each latent embedding vector $\mathbf{e}_i \in \mathbb{R}^D$. The VQ-VAE compresses the input $\mathbf{y}^{(i)}$ to a sequence of discrete latent variables $\mathbf{\tilde{y}}^{(i)}=\{e_0,e_1,...,e_L\} \in \mathbb{N}^{L}_K$. Here, $L = d_1' \times d_2'$ corresponds to the spatial dimensions of the encoded variable $\mathbf{\tilde{y}}^{(i)}$, where $d_1'=d_1/r$ and $d_2'=d_2/r$. The compression ratio $r$ quantifies the reduction in dimensionality achieved during encoding. The decoder then reconstructs the input $\mathbf{y}$ from discrete latent sequences $\mathbf{\tilde{y}}$.  By employing response discretization, any DNA-to-continuous-variable task can be converted into a DNA-to-discrete-token task. 

\begin{table*}[ht!]
\centering
\caption{\textbf{Performance of \method across 8 tasks in the genomic benchmark.} \method (116M) achieves the highest average. $\pm$ indicates the difference between the max and min value in 10 fold cross-validation.}
\resizebox{\textwidth}{!}{
\begin{tabular}{@{}lccccc@{}}
\toprule
\textbf{Task Name} & \textbf{CNN} & \textbf{HYENADNA} & \textbf{CADUCEUS -PH} & \textbf{DNABERT2} & \textbf{Omni-DNA} \\
 & \small\textbf{(264K)} & \small\textbf{(436K)} & \small\textbf{(470K)} & \small\textbf{(117M)} & \small\textbf{(116M)} \\ \midrule
\textbf{MOUSE ENHANCERS}          & $0.715 \pm 0.087$  & $0.780 \pm 0.025$  & $0.754 \pm 0.074$  & $\underline{0.792} \pm 0.031 $  & $\textbf{0.799} \pm 0.004$  \\
\textbf{CODING VS. INTERGENOMIC}  & $0.892 \pm 0.008$  & $0.904 \pm 0.005$  & $0.915 \pm 0.003$  & $\textbf{0.949} \pm 0.002 $    & $\underline{0.942} \pm 0.010$  \\
\textbf{HUMAN VS. WORM}           & $0.942 \pm 0.002$  & $0.964 \pm 0.002$  & $0.973 \pm 0.001$  & $0.975\pm 0.002 $ & $\textbf{0.976} \pm 0.001$ \\
\textbf{HUMAN ENHANCERS COHN}     & $0.702 \pm 0.021$  & $0.729 \pm 0.014$  & $\textbf{0.747} \pm 0.004$  & $0.714 \pm 0.025 $      & $\underline{0.738} \pm 0.002$ \\
\textbf{HUMAN ENHANCER ENSEMBL}   & $0.744 \pm 0.122$  & $0.849 \pm 0.006$  & $\underline{0.893} \pm 0.008$  & $0.891 \pm 0.051 $      & $\textbf{0.919} \pm 0.021$ \\
\textbf{HUMAN REGULATORY}         & $0.872 \pm 0.005$  & $0.869 \pm 0.012$  & $\underline{0.872} \pm 0.011$  & $0.852 \pm  0.024$     & $\textbf{0.895} \pm 0.012$ \\
\textbf{HUMAN OCR ENSEMBL}        & $0.698 \pm 0.013$  & $0.783 \pm 0.007$  & $\textbf{0.828} \pm 0.006$  & $0.789 \pm 0.012$         & $\underline{0.791} \pm 0.001$ \\
\textbf{HUMAN NONTATA PROMOTERS}  & $0.861 \pm 0.009$  & $0.944 \pm 0.002$  & $0.946 \pm 0.007$  & $0.912 \pm 0.013$ & $\textbf{0.968} \pm 0.013$ \\ 
\midrule
\textbf{Average}  & $0.803$  & $0.853$  & $0.866$  & $0.859$ & $\textbf{0.879}$ \\ 
\bottomrule
\end{tabular}
}
\label{tab:gb_reuslts}
\vspace{-1em}
\end{table*}

\section{Results on Conventional Genomics Tasks}
\label{sec:fullfinetune}
\paragraph{Setup} To evaluate the quality of \method and establish a baseline for multi-tasking, we first follow the conventional evaluation process in single task mode. We compare \method with DNABERT-2~\cite{zhou2023dnabert}, NT-Transformer~\cite{dalla2024nucleotide}, HyenaDNA~\cite{nguyen2024hyenadna}, and Caduceus models~\cite{schiff2024caduceus}, across two widely used benchmarks: NT-Downstream Tasks ~\cite{dalla2024nucleotide} and  Genomic Benchmark ~\cite{grevsova2023genomic}. Evaluation is performed using a conventional approach, where a MLP head is attached to the pretrained models, followed by full-size finetuning. The detailed set up of the evaluation is in \cref{app:finetune_ch}.

\subsection{Nucleotide Transformer Downstream Tasks}
\label{sec:nt_classification}

NT Downstream Tasks include 18 distinct binary or three-class classification tasks. These tasks can be categorized into four categories: epigenetic marker~\cite{pokholok2005genome},  promoter~\citep{oubounyt2019deepromoter}, enhancer and splice site prediction. Following the prior evaluation setup~\cite{schiff2024caduceus}, Matthews Correlation Coefficient (MCC) is used for histone marker and enhancer, while F1-score is used for splice and promoter classification. We finetune each model with 10-fold cross-validation \cite{schiff2024caduceus}, with maximum epochs set to 20.

\Cref{tab:pretraining_comparison} shows the average performance of 16 models across different tasks types and the average performance across 18 tasks. Omni-DNA (1B) and Omni-DNA (116M) achieve the best average performance of 0.767 and the second of 0.755. Both models are trained with non-parametric norm. We provide the task-specific performance of 16 models in \Cref{tab:pretraining_comparison_part1}, Omni-DNA achieves superior performance on 13 out of 18 benchmark tasks, surpassing competing methods. Task-specific results in \Cref{tab:pretraining_comparison_part1} show that Omni-DNA outperforms competing methods on 13 out of 18 benchmarks. For the remaining five tasks, Omni-DNA (1B) ranks second and third in Promoter:ALL and Promoter:NonTATA. In the three splice site classification tasks, Omni-DNA models perform lower than NT models but still surpass other models.


\subsection{Genomic Benchmark Results}
Genomic Benchmark (GB)~\cite{grevsova2023genomic} contains eight DNA regulatory element classification tasks, similar to NT downstream tasks, with additional tasks on species classification and gene regulatory element classification on mouse. The number of sequences from seven tasks in GB was 10 times larger compared to NT downstream tasks. The details on the finetuning setting are included in \cref{app:gb_ft}. Each model was finetuned for a max of 10 epochs. We compare the performance of smaller size model from each type of the model in \Cref{tab:gb_reuslts}. 

Omni-DNA (116M) achieved the highest average score, ranking first in five out of eight tasks and second in the remaining three. Compared to DNABERT-2, which has a similar model size, Omni-DNA (116M) outperformed DNABERT-2 in seven out of eight tasks.

\section{Building Cross-modal Multi-task Genomic Foundation Models}

In this section, we demonstrate the \textbf{cross-modal multi-task} capability of Omni-DNA in using DNA sequences to three distinct modalities: discrete labels, textual descriptions, and images. The first two are derived from real-world datasets, while the third is based on a synthetic dataset called \textit{Needle-in-DNA}. Despite the significant differences in task formats, a unified approach is employed to address all three.

For classification problems, Omni-DNA finetuned with multiple tasks not only has the ability to solve more than one task simultaneously but also improves overall model performance on related NT downstream tasks: a phenomenon we refer to as the \textbf{synergistic effect}. Additionally, we extend beyond conventional genomic modeling tasks to tackle more challenging and general problems: (1) DNA-to-function and (2) DNA-to-image. These efforts underscore the potential of training a unified genomic foundation model capable of acting as a generalist across both tasks and modalities.

\subsection{Multi-Task Model for Acetylation and Methylation}

\begin{table*}[t!]
\caption{Performance comparison across 10 related acetylation and methylation tasks. Omni-DNA@mult. outperforms single-task models in 4 tasks, and achieves the second-best performance in 5 tasks, demonstrating the synergistic effect of multi-task genomic fine-tuning.}
\label{tab:multi-tasking}
\resizebox{\textwidth}{!}{
\begin{tabular}{@{}lccccc@{}}
\toprule
          & \small \textbf{Omni-DNA@mult.} & \small \textbf{Omni-DNA@sgl.} & \small\textbf{DNABERT2@sgl.} & \small\textbf{NT Transformer@sgl.} & \small\textbf{CADUCEUS-PH@sgl.} \\ 
          & \small\textbf{(1B)} & \small\textbf{(1B)} & \small\textbf{(117M)} & \small\textbf{(500M)} & \small\textbf{(1.9M)} \\ 
\midrule
H3       & $0.759 \pm 0.021$ & $\mathbf{0.824} \pm 0.032$ & $0.785\pm0.033$ & $0.784\pm0.047$ & $\underline{0.815} \pm 0.048$ \\ 
H3K14AC  & $\mathbf{0.890} \pm 0.011$ & $\underline{0.697} \pm 0.077$ & $0.516\pm0.028$ & $0.551\pm0.021$ & $0.631 \pm 0.026$ \\
H3K36ME3 & $\underline{0.674} \pm 0.008$ & $\mathbf{0.686} \pm 0.002$ & $0.591\pm0.020$ & $0.625\pm0.013$ & $0.601 \pm 0.129$ \\
H3K4ME1  & $\underline{0.565} \pm 0.015$ & $\mathbf{0.617} \pm 0.000$ & $0.511\pm0.028$ & $0.550\pm0.021$ & $0.523 \pm 0.039$ \\
H3K4ME2  & $\mathbf{0.691} \pm 0.008$ & $\underline{0.547} \pm 0.006$ & $0.336\pm0.040$ & $0.319\pm0.045$ & $0.487 \pm 0.170$ \\
H3K4ME3  & $\mathbf{0.785} \pm 0.008$ & $\underline{0.642} \pm 0.001$ & $0.352\pm0.077$ & $0.410\pm0.033$ & $0.544 \pm 0.045$ \\
H3K79ME3 & $\underline{0.709} \pm 0.017$ & $\mathbf{0.752} \pm 0.007$ & $0.613\pm0.030$ & $0.626\pm0.026$ & $0.697 \pm 0.077$ \\
H3K9AC   & $\underline{0.693} \pm 0.013$ & $\mathbf{0.701} \pm 0.002$ & $0.542\pm0.029$ & $0.562\pm0.040$ & $0.622 \pm 0.030$ \\
H4       & $\underline{0.773} \pm 0.013$ & $\mathbf{0.822} \pm 0.005$ & $0.796\pm0.027$ & $0.799 \pm 0.025$ & $0.811 \pm 0.022$ \\
H4AC     & $\mathbf{0.853} \pm 0.014$ & $\underline{0.652} \pm 0.001$ & $0.463\pm0.041$ & $0.495\pm0.032$ & $0.621 \pm 0.054$ \\
\midrule
Average     & $\mathbf{0.739}$ & $\underline{0.694}$ & $0.551$ & $0.572$ & $0.635$ \\
\bottomrule
\end{tabular}
}
\vspace{-1em}
\end{table*}

\paragraph{Task} In the NT downstream tasks, 10 related tasks---H3, H3K14AC, H3K36ME3, H3K4ME1, H3K4ME2, H3K4ME3, H3K79ME3, H3K9AC, H4, and H4AC---are considered. These tasks are interconnected due to the biological correlation between acetylation and methylation effects~\citep{pokholok2005genome}. The objective here is to train a model capable of addressing all 10 tasks with a single fine-tuning step. While conventional fine-tuning approaches with a classification head struggle to address this challenge, \emph{cross-modal multi-task finetuning} unifies these tasks into a single model. At inference time, having an omni-model that solves all tasks eliminates the need for context switching and repeated fine-tuning.

We use Omni-DNA (1B) to complete 10 acetylation and methylation tasks using  multi-task finetuning with NEFTune noise level = 5 and repeating factor = 10. The baselines include the best-performing single-task models: Omni-DNA (1B) and CADUCEUS-PH (1.9M), as presented in \cref{sec:fullfinetune}, along with DNABERT-2 and NT (500M) as references. Conventionally, multi-task finetuning may lead to performance drop compared with single-task finetuning due to the challenge of balancing generalizability across tasks. However, by grouping similar tasks, we observe significant improvements. The results, shown in Table~\ref{tab:multi-tasking} and Figure~\ref{fig:multi-tasking}, demonstrate that Omni-DNA@mult achieves dramatic improvements in 4 tasks, and obtains the highest average accuracy. 
This indicates that the resulting model not only reduces the burden of context switching and repetitive finetuning but also leverages the internal relationships between tasks to achieve higher performance than its single-task counterparts. We refer to this phenomenon as the \textbf{synergistic effect}, which emerges in finetuning genomic foundation models in a multi-task setting.

\subsection{Functional Annotation Generation (DNA2Func)}

\paragraph{Dataset and Task} As shown in \cref{fig:seq2func}, we have constructed a large-scale DNA sequence dataset enriched with text-based functional annotations. 
This dataset comprises 330,000 sequences from various mammalian species, and
each sequence is annotated with a concise short annotation followed by a more detailed natural language description. 
We split the dataset into finetuning and test sets with a 9:1 ratio.
The full workflow for dataset construction is in \cref{app:dna2func}. 
This is a multi-task scenario with two objectives: (1) generating concise short annotations for DNA sequences and (2) further explaining the corresponding functions by natural language.

\paragraph{Baselines} We finetuned our proposed \method, using the finetuning dataset, denoted as Omni-DNA@ft. 
Since conventional genomic foundation models cannot handle these tasks, we employed GPT-4o (GPT4o@zeroshot) to directly predict DNA functions using the prompt in~\cref{tab:prompt-for-dna2func}.
In addition, we included OLMo-1B as a baseline, a similar architecture model pre-trained on natural language, and finetuned on the same finetuning dataset with \method, referred to as OLMo-1B@ft.

\paragraph{Evaluation \& Metrics} To evaluate the accuracy of free-form text is challenging. Thus, we leveraged an LLM (GPT4o) to judge whether the generated text and ground-truth are matched. We first utilized GPT4o to
classify the generated text into seven function labels, as shown in \cref{fig:seq2func}. Then the F1 scores and Matthews Correlation Coefficient (MCC) are computed between the classified labels and ground-truth.




\paragraph{Quantitative Results} As shown in Table~\ref{tab:model_comparison}, Omni-DNA obtains the best result compared to GPT4o and OLMo-1B, highlighting the potential for solving this domain-specific problem. \Cref{tab:dna2func-1,tab:dna2func-2,tab:dna2func-3} present representative responses from the three models. Among them, Omni-DNA@ft consistently demonstrates the strongest performance. In contrast, GPT4o fails to understand the meaning of the DNA sequences in all three examples, whereas the other two models provide more meaningful responses. Notably, Omni-DNA is capable of producing grammatical and novel sentences, rather than merely reproducing funetuning datasets.

\begin{table}[t!]
\centering
\caption{Comparison of Weighted F1 Score and MCC for Omni-DNA@ft, GPT4o@zeroshot, OLMo-1B@ft, and Random Guess.}
\label{tab:model_comparison}
\resizebox{\columnwidth}{!}{
\begin{tabular}{lcccc}
\toprule
 & \textbf{Omni-DNA} & \textbf{GPT4o} & \textbf{OLMo-1B} & \textbf{Random} \\
\midrule
F1 Score     & \textbf{0.730} & 0.659 & 0.701 & 0.483 \\
MCC      & \textbf{0.367} & -0.015 & 0.342 & 0.008 \\
\bottomrule
\end{tabular}
}
\vspace{-1em}
\end{table}

\subsection{Needle-in-DNA Task (DNA2Image)}
 \begin{figure*}[ht!]
    \centering
\begin{scriptsize}
\begin{tabular}{|c|c|cc|cc|cc|cc|}
\hline
\multicolumn{2}{|c|}{Examples From Models} & \multicolumn{2}{c|}{Class 0} & \multicolumn{2}{c|}{Class 1} & \multicolumn{2}{c|}{Class 2} & \multicolumn{2}{c|}{Class 3} \\ \hline
Valid & Invalid & Sequence & Image & Sequence & Image & Sequence & Image & Sequence & Image \\ \hline
\includegraphics[width=0.4cm]{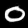} \includegraphics[width=0.4cm]{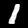} & \includegraphics[width=0.4cm]{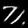} & T...\textcolor{darkgreen}{TATAAA}... & \includegraphics[width=0.4cm]{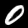} & C...\textcolor{darkgreen}{CAAT}... & \includegraphics[width=0.4cm]{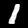} & A...\textcolor{darkgreen}{GGGCGG}... & \includegraphics[width=0.4cm]{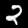} & T...\textcolor{darkgreen}{TTAGGG}... & \includegraphics[width=0.3cm]{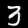} \\ 
\cline{3-10}  
\includegraphics[width=0.4cm]{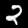} \includegraphics[width=0.4cm]{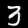} & \includegraphics[width=0.4cm]{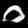} & ...\textcolor{darkgreen}{TATAAA}......A & \includegraphics[width=0.4cm]{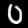} & T...\textcolor{darkgreen}{CAAT}... & \includegraphics[width=0.4cm]{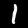} & ...\textcolor{darkgreen}{GGGCGG}...T & \includegraphics[width=0.4cm]{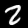} & ...\textcolor{darkgreen}{TTAGGG}...A & \includegraphics[width=0.3cm]{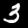} \\ \hline
\end{tabular}
\end{scriptsize}
    \includegraphics[width=0.9\linewidth]{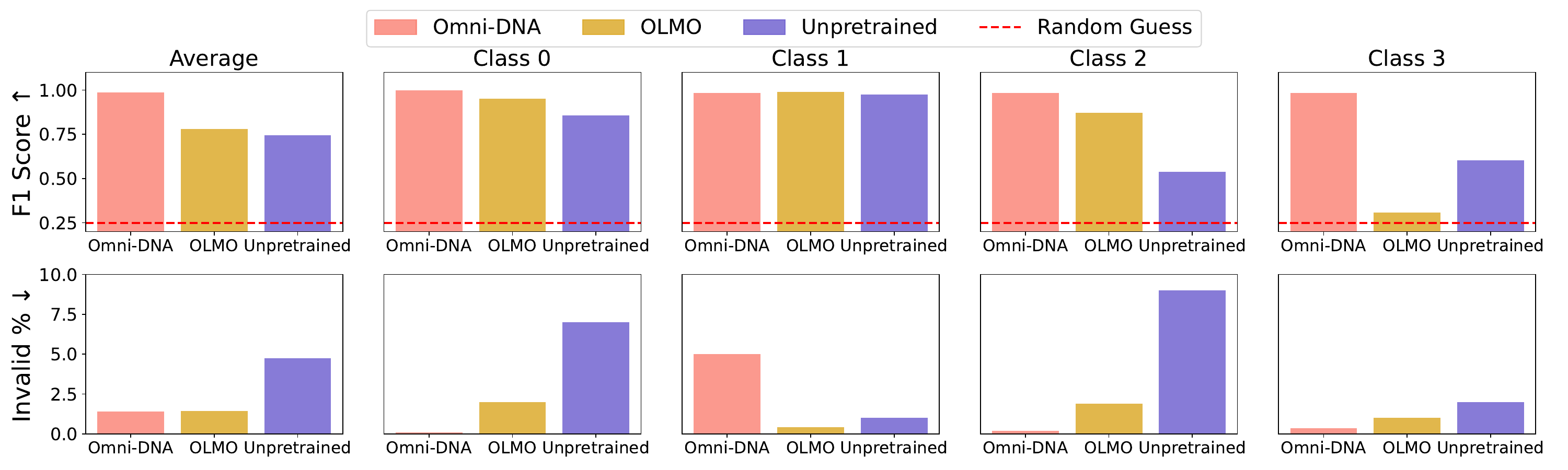}
    \vspace{-1em}
    \caption{\textbf{F1 scores and invalid percentages for \textit{Needle-in-DNA}, averaged and per class}. Omni-DNA outperforms both baselines.}
    \label{fig:dna2iamge}
\vspace{-1em}
\end{figure*}

\paragraph{Dataset and Task} The identification of short functional motifs in DNA is pivotal for understanding gene regulation because these conserved subsequences often serve as transcription factor binding sites or other key regulatory elements~\cite{tompa2005, avsec2021base}. Over the past decade, deep learning methods have further advanced motif discovery by more accurately predicting sequence specificities of DNA- and RNA-binding proteins~\cite{alipanahi2015predicting}, thus refining our ability to pinpoint critical regulatory regions within the genome. Leveraging motif-aware learning for training LLMs, has proven effective for uncovering regulatory and structural elements in genomic sequences ~\cite{wang2024multi, sanabria2024dna}. Motivated by this, we generate a dataset consisting of 48,000 synthetic DNA sequences by a Hidden Markov Model. Each sequence contains one of four functional motifs (referred to as ``needles''): \{TATAAA, CAAT, GGGCGG, TTAGGG\}, and is labeled 0, 1, 2, or 3 accordingly. Notably, the label is represented by an image of a handwritten digit \{\includegraphics[width=0.3cm]{figures/mnist_0.png}, \includegraphics[width=0.3cm]{figures/mnist_1.png}, \includegraphics[width=0.3cm]{figures/mnist_2.png}, \includegraphics[width=0.3cm]{figures/mnist_3.png}\}, sampled from the MNIST dataset~\cite{yadav2019cold}. The dataset is split into finetuning, validation, and test sets in an 8:1:1 ratio.

Following data discretization in \cref{sec:omni-finetuning}, we train a VQ-VAE~\cite{van2017neural} with six quantized vectors of dimension 32 and a compression ratio of 4. This model converts the grayscale images of size $28 \times 28$ into 49 discrete tokens. It leverages a multi-task setting with two tasks: (1) classify the DNA sequences, and (2) generate the corresponding handwritten digit image. The details of dataset construction can be found in \cref{app:needle}.



\paragraph{Evaluation and Baselines} To assess the benefit of pretraining on DNA sequences, we employ two baselines: OLMo-1B, a natural language model pretrained on text with instruction tuning, and the Vanilla Model, which shares \method's architecture but is randomly initialized without pretraining. The baselines and \method (1B) are all fine-tuned on the fine-tuning dataset for 10 epochs.

The evaluation is conducted by two human annotators who independently assess the images generated by the three models. They provide two labels: (1) validity indicating whether the image represents a number from 
\{0,1,2,3\}, and (2) the corresponding digit if valid. 
Inconsistent answers between annotators are discarded. This labeling process identifies two types of errors: vague or non-meaningful images, and incorrect classification. Thus, two metrics are used to measure model performance: (1) \textbf{invalid percentage}, the proportion of generated images that are non-numeric or not in \{0,1,2,3\}, and (2) \textbf{macro F1 scores}, averaged across all valid samples, for each class.


\paragraph{Results} Figure~\ref{fig:dna2iamge} shows that \method achieves a Macro F1 score of 0.987 and an invalid percentage of 1\% on average, significantly outperforming baselines. This indicates that \method nearly solves the task perfectly. In contrast, while OLMo-1B has a high F1 score in the motifs TATAAA (class 0) and CAAT (class 1), it struggles with GGGCGG (class 2) and TTAGGG (class 3). Since its invalid percentages are similar across these classes, the primary issue is generating incorrect digit images, indicating that OLMo-1B fails to classify DNA sequences accurately.


\paragraph{Beyond Memorization} Our visualization results demonstrate that Omni-DNA does not merely memorize the finetuned images, but actually generates novel handwritten digits with various shapes. We verify that these examples are not present in the fine-tuning set, indicating that the model learns general digit patterns in \{0,1,2,3\} conceptually rather than memorizing discretized tokens. See \cref{fig:mnist_sample} for generated examples.

\section{Ablation Study}
\label{sec:ablation}
\subsection{Ablation on Positional Embedding Methods}

\begin{figure}[h!]
    \centering
    \includegraphics[width=\columnwidth]{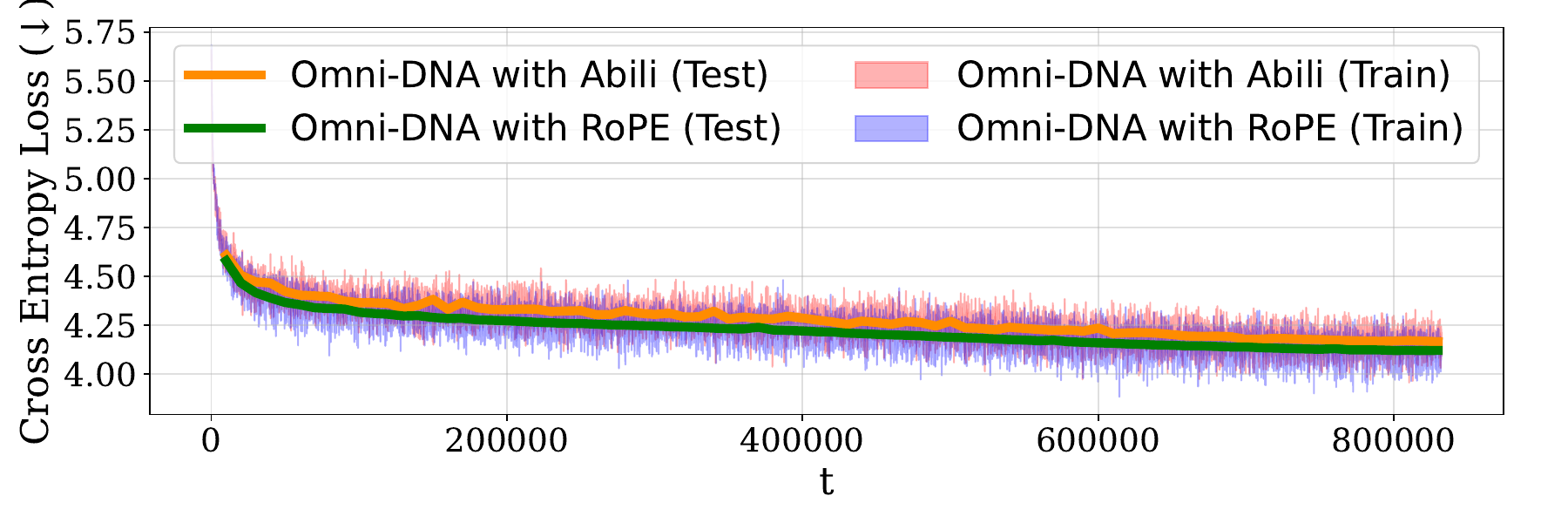}
    \vspace{-2em} 
    \caption{\textbf{Losses for \method(116M) pretrained with ALiBi and RoPE,} indicating RoPE’s faster convergence and lower losses.} 
    \label{fig:embedding}
\end{figure}

We utilized \method(116M) to test the effect of two positional embedding methods: ALiBi~\citep{press2021train} and RoPE~\citep{su2024roformer}. We perform pretraining on 300 billion nucleotides with keeping the remaining hyperparameters unchanged. 
The training and test losses during pretraining are shown in \Cref{fig:embedding}. 
RoPE achieves a lower test loss and converges faster, indicating that it better captures the contextual and relational properties of DNA compared to ALiBi.


\subsection{Larger Models Mitigate Distribution Shifts} 
\begin{figure}[ht!]
    \centering
    \includegraphics[width=\columnwidth]{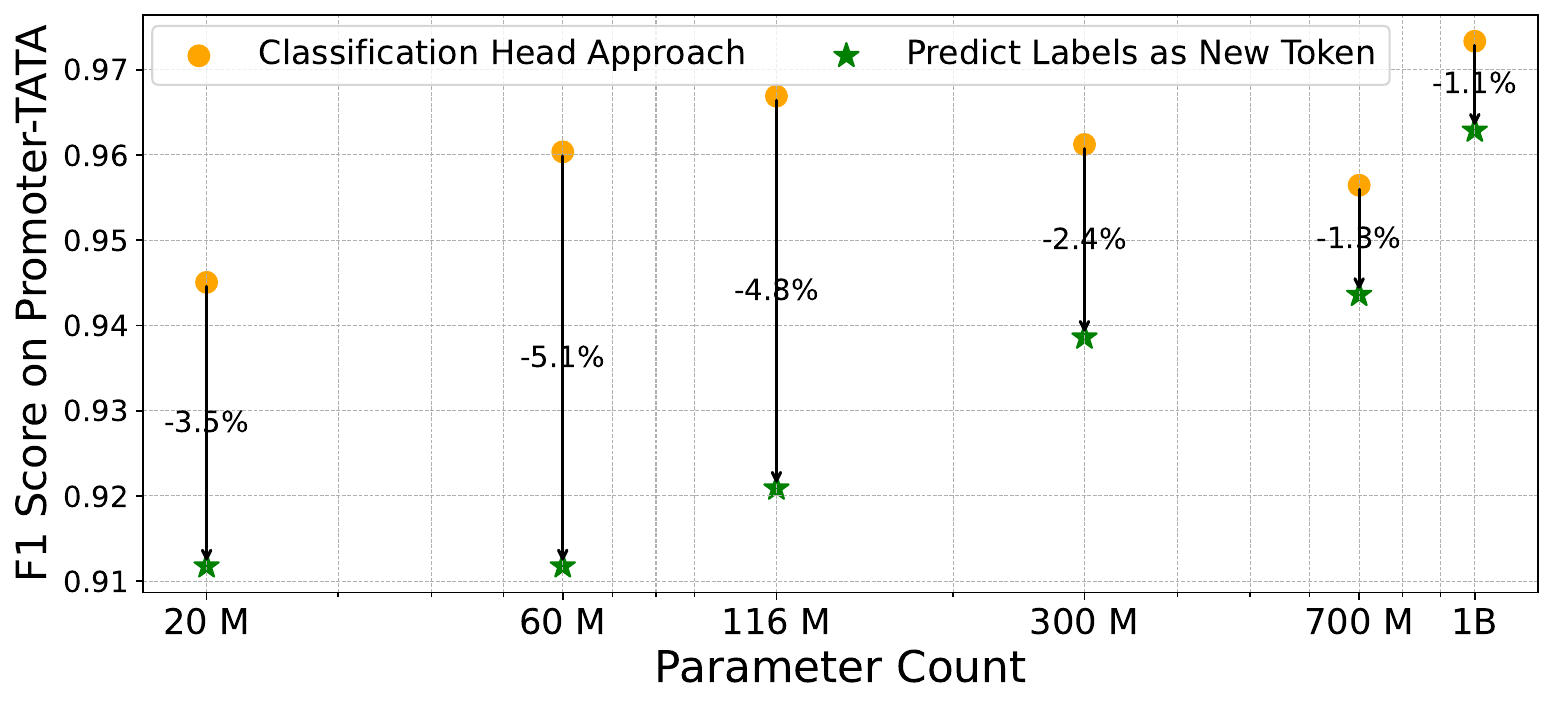}
    \vspace{-1em} 
    \caption{\textbf{Impact of model size on performance degradation due to vocabulary expansion.} Larger models exhibit better resilience against distribution shifts.}
    \label{fig:vocab_expansion}
\end{figure}

\Cref{eq:5} shows that introducing new vocabulary leads to a distribution shift of existing tokens. While synergistic effects in multi-task settings can mitigate this shift \cite{son2024multi}, single-task scenarios experience performance degradation when new tokens are added compared to using a classification head. \Cref{fig:vocab_expansion} shows how performance degradation varies with model size in the promoter TATA classification task. 
Notably, increasing the model size alleviates this issue, and \method(1B) maintains performance comparable to those, prior to vocabulary expansion.

\subsection{Impact of Token Replication Factor}
\begin{figure}[h!]
    \centering
    \includegraphics[width=\columnwidth]{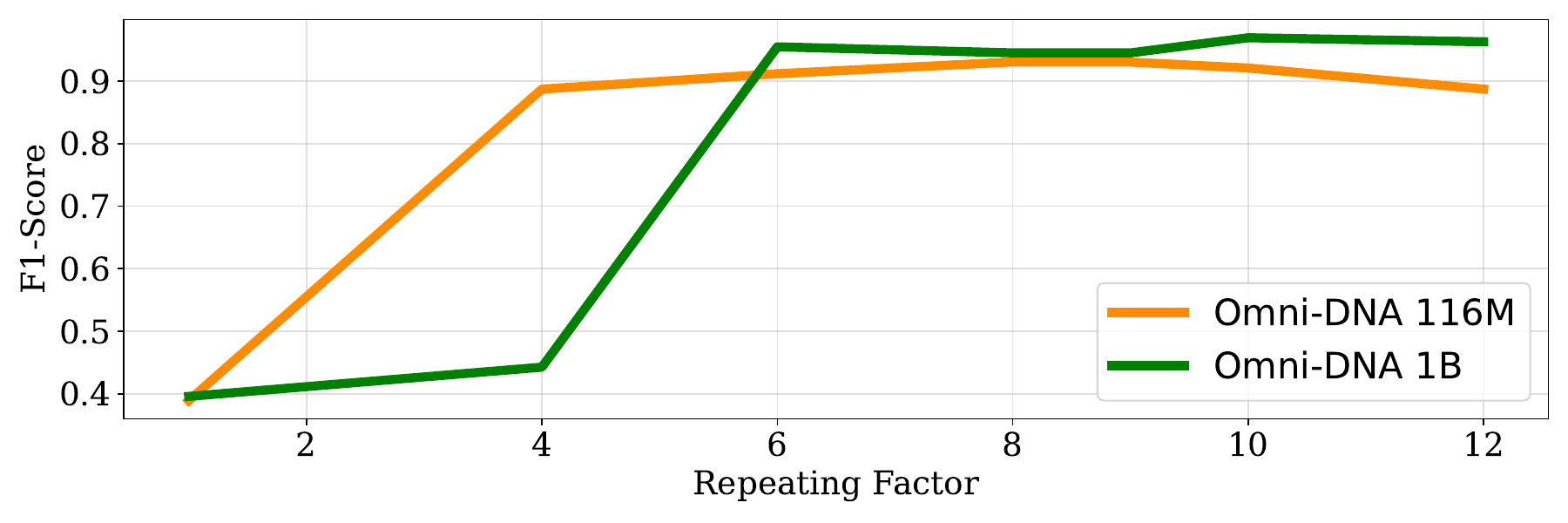}
    \vspace{-2em} 
    \caption{\textbf{Impact of the token replication factor $\alpha$} on the promoter TATA classification task.}
    \label{fig:repearting_factor}
\end{figure}

\Cref{fig:repearting_factor} illustrates the effect of the token replication factor $\alpha$ on the promoter TATA classification task for \method sized 116M and
1B. Without token replication ($\alpha=1$), classification performance is poor. We find that setting $\alpha$ within the range [8, 11] is effective across various task types.

\section{Conclusion}
We have demonstrated that pretraining an auto-regressive transformer on DNA sequences, followed by cross-modal multi-task finetuning, is an effective strategy for constructing a unified genomic foundation model.
This model can handle diverse genomic tasks across multiple modalities, including discrete labels, textual descriptions, and other multi-dimensional outputs.
Specifically, we demonstrate that:
(i) The auto-regressive model can match or outperform the bidirectional transformer on genomic sequence modeling. (ii) The model pretrained solely on genomic sequences can generalize to unseen tokens through finetuning, achieving the SOTA performance. Our proposed \method not only handles  multiple classification tasks within a single model, but also tackles more complex challenges, such as converting DNA into coherent, contextually relevant natural language, and mapping DNA to multi-dimensional representations, as showcased in DNA2Funtion and DNA2image.


We hope that this paper with available codes for pretraining and finetuning, will serve as a starting point for researchers to explore more complex genomic tasks and develop more comprehensive genomics foundation models.


\paragraph{Limitations} Our work primarily focuses on transformer-based architectures, leaving the exploration of alternative approaches, such as state-space models like Mamba \cite{gu2023mamba} and Hyena \cite{nguyen2024hyenadna}, and recurrent neural network architectures like xLSTM \cite{schmidinger2024bio} for future research. Additionally, while our model demonstrates the feasibility of mapping DNA to multi-dimensional outputs, such as structure or genomic assay signals, achieving high-fidelity cross-modality translation remains an open challenge. Future improvements in tokenization strategies \cite{li2024vqdna, pagnoni2024byte, qiao2024model}, including more robust vector quantization (VQ-VAE) techniques, could potentially improve the model’s ability to handle complex modalities with greater precision. Moreover, the DNA pretraining corpus, while extensive at 300B nucleotides, may not fully capture the diversity of regulatory elements across all species and tissue types \cite{andrews2023mammalian}. Finally, while we have demonstrated strong performance on regulatory element classification and selected tasks, our evaluation does not encompass important genomic challenges like long-range interaction prediction \cite{cheng2025dnalongbench} or variant effect prediction \cite{li2024gv}.


\bibliography{example_paper}
\bibliographystyle{icml2025}

\newpage
\appendix
\clearpage
\onecolumn

\section{Pretraining Model Configurations}\label{app:pretrain-config}
The family of Omni-DNA has six models, and their parameters and architectures are shown in Table~\ref{tab:hyperparameters} and~\cref{tab:omni_dna_dnabert_nttransformer}.

\begin{table*}[ht!]
\centering
\caption{Architecture for 6 Omni-DNA models.}
\resizebox{0.5\textwidth}{!}{
\begin{tabular}{@{}ccccc@{}}
\toprule
\textbf{Parameters} & \textbf{Layers} & \textbf{Heads}  & \(\mathbf{d_{\text{model}}}\) & \textbf{LayerNorm} \\ \midrule
20M                & 8              & 8                          & 256             & RMS          \\
60M                & 8              & 8                         & 512             & RMS          \\
116M               & 12             & 12                        & 768             & No Bias      \\
300M               & 16             & 16                        & 1024            & RMS          \\ 
700M               & 16             & 16                        & 1536            & RMS          \\ 
1B                 & 16             & 16                        & 2048            & No Bias      \\ 
\bottomrule
\end{tabular}
\label{tab:hyperparameters}
}
\end{table*}

\begin{table}[ht]
\centering
\caption{Comparison of pretraining setup between Omni-DNA, DNABERT2, and NT-transformer.}
\resizebox{\textwidth}{!}{
\begin{tabular}{lccc}
\toprule
 & \textbf{Omni-DNA (six sizes)} & \textbf{DNABERT2 (116M)} & \textbf{NT-transformer (5 sizes)} \\
\midrule
Layer norm type         & non-parametric/RMSNorm  & parametric        & parametric \\
Positional embeddings   & RoPE            & ALiBi           & RoPE \\
Attention variant       & full            & full            & full \\
Biases                  & none            & in both LN and Attention         & in both LN and Attention \\
Block type              & sequential      & sequential     & sequential \\
Activation             & SwiGLU          & GEGLU         & SwiGLU \\
Sequence length (In Tokens)         & 250            & 128           & 1000 \\
Batch size (Across GPUs)  & 384            & 4096           & 512 \\
Total Steps     & $800000$& $150000$& $16,000$ \\
Warmup steps            & 5000            &  -           & 16,000 \\
Peak LR                 & 3.0E-04         & 5.0E-04        & 1.0E-04 \\
Minimum LR              & 3.0E-05         & 0        & 5.0E-05 \\
Weight decay            & 0.1             & 1e-5            & 0.1 \\
Beta1                   & 0.9             & 0.9            & 0.9 \\
Beta2                   & 0.95            & 0.98           & 0.999 \\
Epsilon                 & 1.0E-05         & 1.0E-05        & 1.0E-08 \\
LR schedule             & linear          & linear         & linear \\
Gradient clipping       & global 1.0      & -    &- \\
Training Precision  & Mxied (bf16 for most operations)        & -           & - \\
Training Framework  & Pytorch Built-in FSDP        &  Pytorch HuggingFace trainer         & - \\
Training Duration (min)   & 8 GPUs* 1 days        &  8GPUs * 14 days        & 128 GPUs*1days \\
Training Duration (max)  & 8  GPUs* 7 days       &  8GPUs * 14 days        &  128 GPUs*28days \\
GPU Types  & A100 Nvidia GPU 40GB       &  Nvidia RTX 2080Ti GPUs.         &  A100 Nvidia GPU \\
\bottomrule
\end{tabular}
}
\label{tab:omni_dna_dnabert_nttransformer}
\end{table}
\section{Task-specific Nucleotide Transformer Benchmark Results}

We present the detailed task-specific results for NT-Downstream tasks in \Cref{tab:pretraining_comparison_part1} and \Cref{tab:pretraining_comparison_part2}. 
\begin{table*}
\caption{Comparison of Pretraining Models and Fine-tuning Performance Across Tasks (Part 1)}
\label{tab:pretraining_comparison_part1}
\resizebox{\textwidth}{!}{
\begin{tabular}{@{}lcccccccccc@{}}
\toprule
\textbf{Metric}           & \textbf{CADUCEUS-PH} & \textbf{CADUCEUS-PS} & \textbf{DNABERT2 [13]} & \textbf{NT50M} & \textbf{NT100M} & \textbf{NT250M} & \textbf{NT500M} & \textbf{NT2.5B} & \textbf{HyenaDNA} \\ \midrule
\textbf{H3}               &       $0.815\pm0.048$              &          $0.799 \pm 0.029$            &        $0.785\pm0.033$                &     $0.745 \pm  0.061$           &      $0.787 \pm  0.010$           &       $0.789 \pm  0.052$          &    $0.784\pm0.047 $             &         $0.814$         &          $0.779\pm0.037$                          \\
\textbf{H3K14AC}          &        $0.631\pm0.026$             &       $0.541\pm0.212$               &         $0.516\pm0.028$               &         $0.471  \pm 0.067$       &     $0.495 \pm   0.101$            &       $0.486 \pm  0.012$          &    $0.551\pm0.021$             &      $0.550$            &            $0.612 \pm0.065$                        \\
\textbf{H3K36ME3}         &          $0.601\pm0.129$           &       $0.609\pm0.109$               &        $0.591\pm0.020$                &      $0.520  \pm 0.054$          &      $0.557  \pm  0.033$           &         $0.571 \pm  0.011$        &    $0.625\pm0.013$             &          $0.632$        &           $0.613 \pm0.041$                         \\
\textbf{H3K4ME1}          &           $0.523 \pm0.039$          &       $0.488\pm0.102$               &      $0.511\pm0.028$                  &        $0.442 \pm  0.021$        &      $0.479 \pm   0.047$           &         $0.486  \pm 0.011$        &    $0.550\pm0.021 $             &      $0.559$            &           $0.512\pm0.024$                         \\
\textbf{H3K4ME2}          &          $0.487\pm0.170$           &          $0.388\pm0.101$            &         $0.336\pm0.040$               &        $0.224  \pm 0.011$        &      $0.261 \pm  0.042$           &      $0.299 \pm   0.017$           &     $0.319\pm0.045 $            &     $0.326$             &            $0.455 \pm0.095$                        \\
\textbf{H3K4ME3}          &           $0.544\pm0.045$          &       $0.440\pm0.202$               &      $0.352\pm0.077$                  &           $0.294 \pm  0.011$     &      $0.336  \pm 0.033$           &      $ 0.360 \pm  0.033$          &       $0.410\pm0.033 $          &        $0.421$          &           $0.549\pm0.056$                         \\
\textbf{H3K79ME3}         &           $0.697\pm0.077 $          &        $0.676 \pm0.026$              &        $0.613\pm0.030$                &         $0.544 \pm 0.016$       &         $0.565\pm   0.089$        &       $0.591 \pm   0.020$          &      $0.626\pm0.026 $           &       $0.642$           &            $0.672\pm0.048$                        \\
\textbf{H3K9AC}           &         $0.622\pm0.030$            &       $0.604 \pm0.048$               &       $0.542\pm0.029 $                 &         $0.489  \pm 0.011$       &      $0.544 \pm 0.021$           &       $0.552  \pm 0.035$          &      $0.562\pm0.040 $           &       $0.575$           &         $0.581\pm0.061$                           \\
\textbf{H4}               &          $0.811\pm0.022$           &        $0.789\pm0.020$              &      $0.796\pm0.027$                  &       $0.784  \pm 0.061$         &         $0.707 \pm  0.060$        &       $0.773 \pm  0.013$          &    $0.799 \pm0.025 $             &       $0.822$           &           $0.763\pm0.044$                         \\
\textbf{H4AC}             &            $0.621\pm0.054$         &          $0.525\pm0.240$            &          $0.463\pm0.041 $              &      $0.417  \pm 0.051$          &        $0.443 \pm  0.008$         &        $0.454 \pm   0.061$         &      $0.495\pm0.032 $           &        $0.501$          &           $0.564 \pm0.038$                         \\
\textbf{Enhancer}   &            $0.546 \pm0.073$         &          $0.491\pm0.066$            &          $0.516\pm0.098$              &          $0.514  \pm 0.004$      &         $0.515 \pm  0.004$        &             $0.519 \pm   0.028$    &        $0.548\pm0.144 $         &         $0.580$         &          $0.517\pm0.117$                          \\
\textbf{Enhancer Types}   &       $0.439\pm0.054$              &      $0.416\pm0.095$                &           $0.423\pm0.051$             &        $0.415  \pm 0.092$        &       $0.435  \pm 0.003$          &      $0.426 \pm   0.031$           &      $0.424 \pm0.132$           &    $0.474$              &        $0.386\pm0.185$                            \\
\textbf{Promoter: ALL}    &         $0.970\pm0.004$            &          $0.967\pm0.004$            &          $0.971 \pm0.006$              &        $0.959  \pm 0.017$        &      $0.967 \pm 0.132$           &        $0.972 \pm   0.008$         &     $0.976\pm0.006 $            &       $0.974$           &         $0.960\pm0.005$                           \\
\textbf{Promoter: NONTATA}&       $0.969\pm0.011$              &        $0.968\pm0.006$              &           $0.972 \pm0.005$             &      $0.956 \pm   0.010$          &         $0.967 \pm 0.045$        &        $0.973 \pm   0.010$         &      $0.976\pm0.005 $           &      $0.977$            &         $0.959\pm0.008$                           \\
\textbf{Promoter: TATA}   &        $0.953\pm0.016$             &       $0.957\pm0.015$               &            $0.955\pm0.021$            &       $0.946  \pm 0.041$         &        $0.937 \pm 0.032$         &        $0.960 \pm 0.03$         &       $0.966\pm0.013 $          &       $0.964$           &            $0.944\pm0.040$                        \\
\textbf{Splice All}       &        $0.940\pm0.027$             &        $0.927\pm0.021$              &           $0.939\pm0.009$             &      $0.978  \pm 0.016$          &     $ 0.986 \pm 0.090$           &        $0.979 \pm  0.071$         &       $0.983\pm0.008 $          &       $0.983$           &          $0.956 \pm0.011$                          \\
\textbf{Splice Acceptor}  &         $0.937\pm0.033$            &        $0.936\pm0.077$              &        $0.975 \pm0.006$                &       $0.981  \pm 0.004$         &      $ 0.979  \pm 0.033$           &      $0.985 \pm  0.068$           &        $0.981\pm0.011 $         &      $0.990$            &         $0.958\pm0.010$                           \\
\textbf{Splice Donor}     &         $0.948\pm0.025$            &         $0.874\pm0.289$             &       $0.963 \pm0.006$                 &       $0.981  \pm 0.078$         &        $0.983 \pm  0.043$         &       $0.985 \pm  0.091$          &       $0.985\pm0.022 $          &       $0.984$           &        $0.949\pm0.024$                            \\
\textbf{\#Params}         & 1.9M                & 1.9M                 & 150M                   & 50M            & 100M            & 250M            & 500M            & 2.5B             & 1.6M                              \\

\bottomrule
\end{tabular}
}
\end{table*}

\begin{table*}
\caption{Comparison of Pretraining Models and Fine-tuning Performance Across Tasks (Part 2)}
\label{tab:pretraining_comparison_part2}
\resizebox{\textwidth}{!}{
\begin{tabular}{@{}lccccccc@{}}
\toprule
\textbf{Metric}           & \textbf{OLMo-DNA20M} & \textbf{OLMo-DNA60M} & \textbf{OLMo-DNA116M} & \textbf{OLMo-DNA300M} & \textbf{OLMo-DNA700M} & \textbf{OLMo-DNA1B} \\ \midrule
\textbf{H3}               &      $0.778 \pm 0.022$                &         $0.775 \pm  0.022$              &        $0.818\pm0.005$                &        $0.820 \pm 0.005$               &      $0.813\pm 0.019$                 &      $0.824 \pm 0.032$      \\
\textbf{H3K14AC}          &        $0.514 \pm 0.043$              &           $0.566 \pm  0.050$            &        $0.685\pm 0.014$                &         $0.639 \pm 0.010$              &      $0.672\pm 0.031 $                 &          $0.697 \pm 0.077$          \\
\textbf{H3K36ME3}         &          $0.525\pm0.16$            &            $0.525+0.019$           &                $0.661\pm 0.013$        &               $0.648 \pm 0.010$        &          $0.689\pm 0.096 $             &           $ 0.686 \pm 0.002$          \\
\textbf{H3K4ME1}          &          $0.393\pm 0.016$            &         $0.400\pm0.085$              &            $0.577\pm 0.083$            &          $0.543 \pm 0.076$             &         $0.577\pm 0.101$              &           $ 0.617 \pm 0.000$          \\
\textbf{H3K4ME2}          &           $0.382\pm 0.110$           &          $0.407 +0.041$             &             $0.576\pm 0.003$           &             $0.515\pm 0.007$          &         $0.568\pm 0.025$              &            $0.547\pm 0.006$          \\
\textbf{H3K4ME3}          &            $0.435\pm 0.190$          &            $0.298  \pm 0.181$           &          $0.587\pm 0.222$              &       $0.628\pm 0.120$                &      $0.584\pm 0.076$                 &         $0.642\pm 0.001$            \\
\textbf{H3K79ME3}         &           $0.588 \pm0.008$           &        $0.655\pm 0.011$               &           $0.718\pm 0.027$             &          $0.707\pm 0.028$             &      $0.730\pm 0.087$                 &         $ 0.752 \pm 0.007$         \\
\textbf{H3K9AC}           &           $ 0.525 \pm0.027$           &           $0.559 \pm0.023$            &          $0.658\pm 0.029$              &         $0.652\pm 0.031$              &        $0.663\pm 0.051$               &         $ 0.701 \pm 0.002$            \\
\textbf{H4}               &            $0.756 \pm0.035$          &         $0.795+ 0.091$              &             $0.802\pm 0.002$           &            $0.786\pm 0.014$           &         $0.793\pm 0.045$              &         $ 0.822 \pm 0.005$             \\
\textbf{H4AC}             &            $0.485\pm 0.059$          &         $0.538 \pm  0.047$              &         $0.663\pm 0.029$               &        $0.659\pm 0.019$               &      $0.656\pm 0.014$                 &          $  0.652 \pm 0.001$          \\
\textbf{Enhancer }        &          $0.511 \pm 0.049$            &           $0.558 \pm 0.074 $            &        $0.593\pm 0.005$                &        $0.545\pm 0.006$               &     $0.596\pm 0.010$                  &           $ 0.580 \pm 0.018$         \\
\textbf{Enhancer Types}   &          $0.457\pm0.014$            &             $0.511\pm0.117$          &             $0.498\pm 0.001$           &           $0.426\pm 0.004$            &        $0.443\pm 0.062$               &           $0.492   \pm 0.023$        \\

\textbf{Promoter: ALL}    &            $0.962 \pm  0.001$          &           $0.962\pm   0.008$            &       $0.973\pm 0.002$                 &       $0.972\pm 0.003$                &      $0.971\pm 0.006$                 &            $0.973 \pm 0.001$         \\
\textbf{Promoter: NONTATA}&            $0.964 \pm 0.006$          &           $0.964 \pm 0.007$          &           $0.972\pm 0.016$             &         $0.973\pm 0.021$              &     $0.975\pm 0.005$                  &           $ 0.975 \pm 0.001$          \\
\textbf{Promoter: TATA}   &           $0.945+0.016$           &            $0.945 +0.017$           &                $0.967\pm 0.001$        &             $0.961\pm 0.027$          &         $0.956\pm 0.016$              &           $0.973 \pm 0.002$           \\
\textbf{Splice All}       &           $0.835 \pm 0.145$           &          $0.835 \pm 0.014$             &         $0.927\pm 0.025$               &        $0.940\pm 0.016$               &      $0.948\pm 0.026$                 &         $0.941 \pm 0.003$             \\
\textbf{Splice Acceptor}  &             $0.931\pm 0.001$         &          $0.930 \pm 0.003$             &          $0.968\pm 0.002$              &         $0.970\pm 0.009$              &       $0.968\pm 0.018$                &         $  0.972 \pm 0.001 $           \\
\textbf{Splice Donor}     &            $0.927+0.023$          &           $0.927 \pm 0.023$            &            $0.951\pm 0.007$            &         $0.957 \pm 0.091$              &         $0.961\pm 0.031$              &           $ 0.963 \pm 0.001$          \\
\textbf{\#Params}         & 20M                  & 60M                   & 116M                   & 300M                  & 700M                  & 1B                  \\ 
\bottomrule
\end{tabular}
}
\end{table*}


\section{Detailed Deduplication Process}
\label{app:deduplication}
\subsection{Handling Non-\{A, T, G, C\} Characters}
The raw genome data from the reference genome includes additional characters beyond \{A, T, G, C\}, which represent ambiguities or gaps. These characters are defined as follows:
\begin{itemize}
    \item \textbf{N}: Represents any nucleotide (A, T, G, or C).
    \item \textbf{R}: Represents purines (A or G).
    \item \textbf{Y}: Represents pyrimidines (C or T).
    \item Other characters (e.g., \textbf{W}, \textbf{S}, \textbf{K}, \textbf{M}, etc.) represent specific nucleotide subsets or unknown bases.
\end{itemize}
These characters were removed during preprocessing to retain only the core nucleotide sequences (\{A, T, G, C\}), ensuring consistency and facilitating the deduplication process.

\subsection{Chunk-Based Deduplication}
The deduplication procedure was performed as follows:
\begin{enumerate}
    \item \textbf{Chunking}: The genome was divided into non-overlapping chunks of 1024 base pairs.
    \item \textbf{Exact Matching}: Identical sequences across the dataset were identified and removed.
    \item \textbf{Efficiency}: This step utilized hashing techniques and optimized string comparison algorithms to handle the large dataset efficiently.
\end{enumerate}

\subsection{Results}
The raw genome dataset initially contained approximately 170 billion nucleotides. After applying the deduplication process, the dataset was reduced to 30 billion nucleotides, representing unique sequences across multiple species.

\subsection{Rationale for Multiple Epochs}
Although deduplication significantly reduced the size of the dataset, the smaller dataset was iterated over multiple epochs during pretraining. This approach ensured that repeated sequences were temporally separated during the training process. By maximizing the temporal distance between occurrences of the same sequence, the risk of overfitting was mitigated, leading to better generalization performance.
\section{Finetuning with Classificatoin Head}
\label{app:finetune_ch}
\label{app:nt_downstream_ft}
\label{app:gb_ft}
\paragraph{Basic Setup} 
We closely follow the setup from Caduceus~\cite{schiff2024caduceus} for evaluating Omni-DNA. The following models are compared: \method, DNABERT-2~\cite{zhou2023dnabert}, NT-Transformer~\cite{dalla2024nucleotide}, HyenaDNA~\cite{nguyen2024hyenadna}, and the Caduceus models~\cite{schiff2024caduceus}.  

For NT Downstream tasks, we use a maximum fine-tuning epoch of 20, while for the Genomic Benchmark (GB) tasks, we use a maximum of 10 epochs. Both NT Downstream and GB tasks include a training set and a test set. For hyperparameter search, 10\% of the training set is reserved as a validation set.  

\paragraph{Hardware \& Framework} 
All fine-tuning is conducted on a single NVIDIA A100 40GB GPU. We utilize the Hugging Face \texttt{Trainer} API for full-size fine-tuning of the pretrained checkpoints. Models are loaded using the \texttt{AutoModelForSequenceClassification} class, which automatically adds a linear layer on top for sequence classification.  

Our experimental results align with those reported in Caduceus~\cite{schiff2024caduceus}. For consistency, we report statistics from~\cite{schiff2024caduceus} for two Caduceus models, DNABERT-2, NT-Transformer 500M, and HyenaDNA in NT Downstream tasks, as well as CNN, HyenaDNA, and CADUCEUS-PH in GB tasks. For NT2.5B, we use the results from its original paper~\cite{dalla2024nucleotide}.

We replicate experiments for the following models: NT50M, NT100M, NT250M, along with six \method models on 18 NT Downstream tasks. Additionally, we fine-tune DNABERT-2 and \method 116M on the eight Genomic Benchmark tasks.  

\paragraph{Hyperparameters} 
A simple grid search is performed for each model using hyperparameters provided in \cref{tab:finetuning_hyperparameters}.  
\begin{table}[ht]
\centering
\caption{Finetuning hyperparameters.}
\resizebox{0.5\textwidth}{!}{
\begin{tabular}{lc}
\toprule
\textbf{Hyperparameter} & \textbf{Value} \\
\midrule
Peak learning rate & $[3 \times 10^{-4}, 1 \times 10^{-5}, 5 \times 10^{-6} ]$ \\
Per device train batch size & [8,16,32] \\
Max gradient norm & 1.0 \\
Weight decay & 0.1 \\
Adam $\beta_1$ & 0.9 \\
Adam $\beta_2$ & 0.999 \\
Adam $\epsilon$ & $1 \times 10^{-8}$ \\
Optimizer & ADAMW \\
\bottomrule
\end{tabular}
\label{tab:finetuning_hyperparameters}
}
\end{table}

\paragraph{Additional Notes}
We observe a performance gap in HyenaDNA between our replication results, which align with those of~\cite{schiff2024caduceus}, and the results reported in the original paper~\cite{nguyen2024hyenadna}, both on NT downstream and GB tasks. Our replication follows our own setup, yielding results consistent with~\cite{schiff2024caduceus}. Upon reviewing the code within the container image provided by~\cite{nguyen2024hyenadna}, we identified additional techniques such as Exponential Moving Average (EMA) and data augmentation that were employed, which may account for the discrepancy. In this work, we use HyenaDNA results obtained without these techniques as the baseline.
\section{Needles in a DNA}
\label{app:needle}

For random DNA sequence generation, we use the Expectation-Maximization (EM) algorithm implemented in the Python package \texttt{hmmlearn} to train a Hidden Markov Model (HMM) with two states, learning both the transition and emission probabilities. We then insert one of the functional motifs from \{TATAAA, CAAT, GGGCGG, TTAGGG\} into the randomly generated sequences. Finally, we filter the sequences, preserving only those that contain exactly one instance of a motif. We generate 12,000 sequences for each motif type.

Next, we process the MNIST images through a trained Vector Quantized Variational Autoencoder (VQ-VAE) to obtain discretized token representations corresponding to each motif type. The pretraining and architecture details of the VQ-VAE are described in the following section.

\subsection{VQ-VAE}

To learn discrete representations of image features, we employ a Vector Quantized Variational Autoencoder (VQ-VAE), which consists of an encoder, a vector quantization layer, and a decoder.

\subsubsection{Encoder}
The encoder transforms input images into a latent space representation. It is implemented as a convolutional neural network (CNN) with residual connections. The architecture consists of:

\begin{itemize}
    \item An initial $4 \times 4$ convolution with stride 2 and padding 1, reducing the spatial dimensions while increasing feature depth.
    \item A second $4 \times 4$ convolution with stride 2, further downsampling the input.
    \item A $3 \times 3$ convolution maintaining the number of hidden channels.
    \item A residual stack of multiple layers with $3 \times 3$ and $1 \times 1$ convolutions, enhancing feature extraction.
\end{itemize}

The output of the encoder is then passed through a $1 \times 1$ convolution to adjust the feature dimensionality before quantization.

\subsubsection{Vector Quantization}
The latent representations are discretized using vector quantization, which maps continuous encodings to the nearest prototype in a learned codebook. We use two different approaches: 

\begin{itemize}
    \item The standard VQ-VAE, where the codebook is learned via backpropagation.
    \item The exponential moving average (EMA) variant, which updates the codebook using an EMA of past embeddings to stabilize training.
\end{itemize}

Given a latent representation $z_e$, the closest embedding vector $z_q$ is selected based on the Euclidean distance. The loss function includes a commitment loss term, weighted by a hyperparameter $\beta$, ensuring that encoded representations stay close to their assigned embedding vectors.

\subsubsection{Decoder}
The decoder reconstructs the input from the discretized representation. It mirrors the encoder with transposed convolutions for upsampling:

\begin{itemize}
    \item A $3 \times 3$ convolution to process the quantized representation.
    \item A residual stack for improved reconstruction.
    \item Two transposed convolutions ($4 \times 4$, stride 2) to upsample back to the original image size.
\end{itemize}

\subsubsection{Training and Hyperparameters}
We train the VQ-VAE using the Adam optimizer with a learning rate of $5 \times 10^{-4}$. The following hyperparameters are used:

\begin{itemize}
    \item Number of hidden channels: 128
    \item Number of residual layers: 2
    \item Number of residual hidden units: 32
    \item Codebook size: 6
    \item Embedding dimension: 32
    \item Commitment cost ($\beta$): 0.25
    \item Decay factor (for EMA variant): 0.99
\end{itemize}

The reconstruction loss is measured using mean squared error (MSE), and training is monitored with additional metrics such as perplexity and codebook usage. The trained VQ-VAE is used to convert MNIST images into discrete tokens, which are then linked to DNA motifs for downstream analysis.

 \begin{figure*}[ht!]
    \centering
    \includegraphics[width=0.5\linewidth]{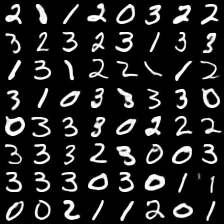}
    \caption{\textbf{Images Generated by Omni-DNA.} These digits are novel in the sense that they do not exist in the training set.}
    \label{fig:mnist_sample}
\end{figure*}

\newpage
\section{Functional Annotation Generation}
\label{app:dna2func}
 \begin{figure*}[ht!]
    \centering
    \includegraphics[width=0.93\linewidth]{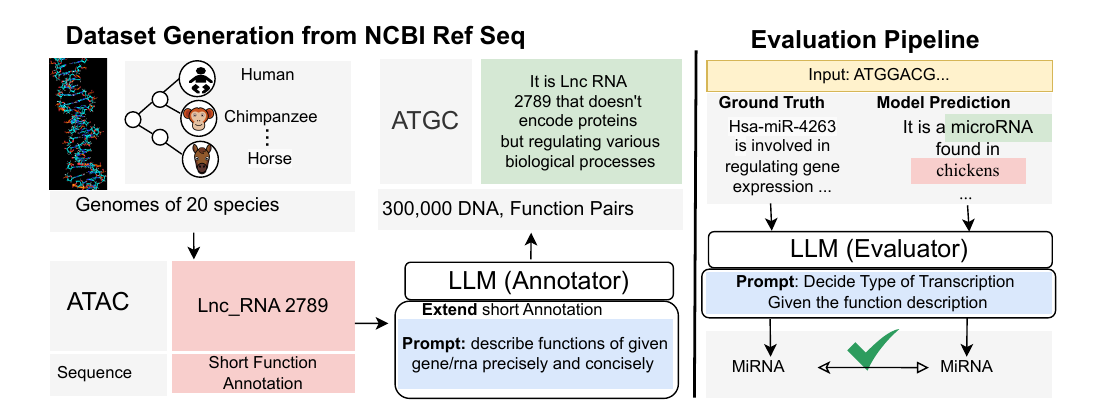}
    \vspace{-2em} 
    \caption{\textbf{Seq2Func Dataset Construction and Evaluation Pipeline.} Genomic sequences from 20 species are annotated with functional descriptions using an LLM, generating 300,000 DNA-function pairs. The evaluation pipeline compares model predictions against ground truth, with an LLM determining transcription type accuracy.}
    \label{fig:seq2func}
\end{figure*}

\paragraph{Dataset Construction Process}

The Seq2Func dataset is constructed using genomic sequences sourced from the \textbf{NCBI Reference Sequence (RefSeq)} database, incorporating sequences from \textbf{20 different species}. The selected species and their genome assemblies include:

\begin{itemize}
\item GCF-000001405.40, human
\item GCF-000001635.27, mouse
\item GCF-036323735.1, rat
\item GCF-028858775.2, Chimpanzee
\item GCF-018350175.1, cat
\item GCF-011100685.1, dog
\item GCF-000696695.1, burgud
\item GCF-000003025.6, pig
\item GCF-016772045.2, sheep
\item GCF-016699485.2, chicken
\item GCF-025200985.1, fly
\item GCF-000002035.6, Zebrafish
\item GCF-029289425.2, Pygmy Chimpanzee
\item GCF-029281585.2, Western gorilla
\item GCF-028885655.2, Pongo abelii
\item GCF-028885625.2, Bornean orangutan
\item GCF-003339765.1, Rhesus monkey
\item GCF-037993035.1, Macaca fascicularis
\item GCF-003668045.3, Chinese hamster
\item GCF-041296265.1, horse
\end{itemize}

The construction process follows these key steps:

\paragraph{Sequence Extraction and Functional Annotation}

\begin{itemize}
\item The raw dataset includes a diverse set of RNA sequences with different functional annotations.
\item The original dataset consists of the following RNA types and their counts:
\begin{itemize}
\item mRNA: 220397
\item tRNA: 20674
\item snRNA: 18996
\item snoRNA: 15577
\item lncRNA: 14291
\item miRNA: 12238
\item primary-transcript: 8644
\item rRNA: 6225
\item transcript: 3860
\item ncRNA: 477
\item guide-RNA: 143
\item antisense-RNA: 28
\item RNase-P-RNA: 9
\item V-gene-segment: 7
\item scRNA: 7
\item Y-RNA: 6
\item telomerase-RNA: 5
\item vault-RNA: 4
\item SRP-RNA: 4
\item RNase-MRP-RNA: 3
\item C-gene-segment: 2
\item D-gene-segment: 1
\end{itemize}
\item To refine the dataset, only \textbf{seven functional RNA types} are retained:
\begin{itemize}
\item mRNA
\item tRNA
\item snRNA
\item snoRNA
\item lncRNA
\item miRNA
\item rRNA
\end{itemize}
\end{itemize}

\paragraph{Annotation Enhancement Using a Large Language Model (LLM)}

\begin{itemize}
\item A \textbf{LLM Annotator} is employed to extend and refine functional annotations.
\item The model is prompted with: \textit{Describe the functions of the given gene/RNA precisely and concisely.''}
    \item For example, an initial annotation like Lnc RNA 2789'' is expanded into:
\begin{quote}
``It is Lnc RNA 2789 that doesn't encode proteins but regulates various biological processes.''
\end{quote}
\end{itemize}

\paragraph{Final Dataset Composition}

\begin{itemize}
\item The final dataset consists of \textbf{300,000 DNA-function pairs}, where each DNA sequence is paired with an enhanced functional description.
\item These functionally annotated sequences serve as high-quality input for machine learning models.
\end{itemize}

\begin{table}[h]
    \centering
    \renewcommand{\arraystretch}{1.5}
    \begin{tabular}{|p{5cm}|p{10cm}|}
        \hline
        \textbf{Usage}  & \textbf{Prompt} \\
        \hline
          Used  by \textbf{LLM (Annotator)} for extending the short annotation to detailed function annotation& 
          \textbf{(System Prompt)} You are a helpful assistant that answers functions of given gene/rna precisely and concisely.
         \\
        \hline
        Used by \textbf{LLM (Evaluator)} for deciding the type of DNA given its function description & 
        \textbf{(System Prompt)} You are a helpful assistant that determines the type of RNA based on the given function description. When deciding the functoin. Your answer should to only be one of the ['mRNA', 'tRNA', 'snRNA', 'snoRNA', 'lnc\-RNA', 'miRNA', 'rRNA'] \\
        \hline
        Given to \textbf{GPT4o@zeroshot} for performing zeroshot DNA function prediction & 
        \textbf{(System Prompt)} You are a helpful assistant that determines the type of RNA based on the given DNA sequence. Your answer should only be one of the mRNA, tRNA, snRNA, snoRNA, lnc\-RNA, miRNA, rRNA\\
        \hline
       
    \end{tabular}
    \caption{Prompt Used during dataset construction, evaluation and zeroshot prediction}
    \label{tab:prompt-for-dna2func}
\end{table}

\begin{table}[h]
    \centering
    \renewcommand{\arraystretch}{1.5}
    \begin{tabular}{|p{4cm}|p{4cm}|p{4cm}|p{4cm}|}
        \hline
        \textbf{Ground Truth} & \textbf{GPT4o} & \textbf{Olmo@ft} & \textbf{OmniDNA@ft} \\
        \hline
        \multicolumn{4}{|p{16cm}|}{\textbf{(Input)} TGGCAGAGATATG...AAGAAAAAAGAAAATGAACTTGGTGCAGGCA (DNA of length 1000) } \\
        \hline
       mRNA:CD209, also known as DC-SIGN (Dendritic Cell Specific Intercellular adhesion molecule-3-grabbing Nonintegrin), is a type I transmembrane protein primarily expressed on dendritic cells and some other immune cells. It functions as a receptor that binds and internalizes pathogens, including viruses and bacteria, thereby facilitating their presentation to  \colorbox{lightgrey}{T cells} and playing a critical role in the \colorbox{lightgrey}{immune response}. CD209 is involved in the recognition of mannose-rich ligands, mediating cellular adhesion, and influencing various aspects of \colorbox{lightgrey}{immune activation} and tolerance. Additionally, it can \colorbox{lightgrey}{modulate inflammatory} responses and has been implicated in HIV-1 and other pathogen infections. & The provided DNA sequence likely represents a genomic region involved in regulatory or coding functions. Based on its composition, it may contain promoter elements, coding regions, untranslated regions (UTRs), or repetitive sequences.  & The mRNA for the putative POM121-like protein 1-like likely encodes a protein that is similar to POM121, which is a component of the nuclear pore complex. This protein may play a role in nucleocytoplasmic transport, facilitating the movement of molecules between the nucleus and cytoplasm. The "putative" designation suggests that its function is inferred from homology to other known proteins, but it may not yet be fully characterized in terms of its specific biological roles or mechanisms. & C-C motif chemokine ligand 22 ( CCL22 ) is a chemokine involved in immune responses . It primarily attracts and \colorbox{lightgreenRGB}{activates T cells} , particularly regulatory T cells ( Tregs ), and plays a role in modulating inflammation and \colorbox{lightgreenRGB}{immune responses}. CCL22 is produced by various cell types , including macrophages and dendritic cells , and is implicated in various diseases , including autoimmune disorders and cancer. \\
        \hline
    \end{tabular}
    \caption{Comparison of Ground Truth, GPT4o, Olmo@ft, and OmniDNA@ft in RNA type determination (Example 1)}
    \label{tab:dna2func-1}
\end{table}

\begin{table}[h]
    \centering
    \renewcommand{\arraystretch}{1.5}
    \begin{tabular}{|p{4cm}|p{4cm}|p{4cm}|p{4cm}|}
        \hline
        \textbf{Ground Truth} & \textbf{GPT4o} & \textbf{Olmo@ft} & \textbf{OmniDNA@ft} \\
        \hline
\multicolumn{4}{|p{16cm}|}{\textbf{(Input)} GAGGTAGTAGATTGTATAG} \\
        \hline
       \colorbox{lightgrey}{miR}-3596d is a microRNA found in rats (Rattus norvegicus). It is involved in the regulation of gene expression at the post-transcriptional level, primarily by binding to complementary sequences in messenger RNAs (mRNAs), leading to their \colorbox{lightgrey}{degradation} or \colorbox{lightgrey}{inhibition} of \colorbox{lightgrey}{translation}. While specific target genes may vary, miRNAs like rno-miR-3596d generally play roles in various biological processes such as development, cell differentiation, proliferation, and apoptosis. Specific roles or targets for rno-miR-3596d would require experimental validation or further studies to elucidate its functions in rat biology. & The DNA sequence GAGGTAGTAGATTGTATAG could serve various functions depending on its genomic context. Generally, it could be:
	1.	A regulatory element – If found in promoter or enhancer regions, it may influence gene expression.
	2.	A binding site – It could be a recognition site for transcription factors, such as ETS-family proteins (which recognize GGAA/T motifs).
	3.	A coding or non-coding region – If within a gene, it might encode a peptide or function as a non-coding RNA.
	4.	A structural or repeat element – It could contribute to chromatin organization or genome stability.For precise functional annotation, its location within a genome and experimental validation would be needed. & hsa-miR-4684 is a microRNA (miRNA) found in humans. It is involved in the regulation of gene expression by binding to complementary sequences in target messenger RNAs (mRNAs), leading to their degradation or inhibition of translation. miR-4684 may play roles in various biological processes, including development, cell proliferation, and differentiation. Specific target genes and pathways regulated by hsa-miR-4684 may vary, and ongoing research aims to elucidate its precise functions & gga - miR - 65 77 - 5p is a microRNA ( miRNA ) found in chickens ( Gallus gallus ). MicroRNAs are small , non - coding RNA molecules that play a crucial role in regulating gene expression by binding to complementary sequences on target messenger RNAs ( mRNAs ), leading to their \colorbox{lightgreenRGB}{degradation} or \colorbox{lightgreenRGB}{inhibition} of \colorbox{lightgreenRGB}{translation} . Specifically , gga - miR - 65 77 - 5p may be involved in various biological processes , including development , differentiation , and response to stress , although specific target genes and functions may require further study \\
        \hline
    \end{tabular}
    \caption{Comparison of Ground Truth, GPT4o, Olmo@ft, and OmniDNA@ft in RNA type determination (Example 2)}
    \label{tab:dna2func-2}
\end{table}

\begin{table}[h]
    \centering
    \renewcommand{\arraystretch}{1.5}
    \begin{tabular}{|p{4cm}|p{4cm}|p{4cm}|p{4cm}|}
        \hline
        \textbf{Ground Truth} & \textbf{GPT4o} & \textbf{Olmo@ft} & \textbf{OmniDNA@ft} \\
        \hline
\multicolumn{4}{|p{16cm}|}{\textbf{(Input)} 
TGCTCACTTCAGCAGCACATGTGCTAAAAT
TGGAATGACACAAAGAAGATTAGCATGGCCGC
TGAGCAAGGATGACACGCACATTTGTGAATTGTTCCATAAAAAA} \\
        \hline
       nRNA:U6 spliceosomal RNA is a small nuclear RNA (snRNA) that plays an essential role in the splicing of pre-mRNA in eukaryotic cells. It is a core component of the spliceosome, the complex responsible for the removal of introns from pre-mRNA transcripts. U6 engages in base-pairing interactions with other spliceosomal RNAs and pre-mRNA, facilitating the assembly and function of the spliceosome. Its primary roles include catalyzing the cleavage of introns and the ligation of exons during the splicing process. U6 is also involved in the activation of the catalytic core of the spliceosome, making it crucial for proper gene expression. & The given DNA sequence likely encodes a functional genetic element, such as a protein-coding region or a regulatory sequence. To determine its precise function, I will need to analyze it for potential coding regions, motifs, or regulatory elements. & U6 spliceosomal RNA is a small nuclear RNA (snRNA) that plays a crucial role in the splicing of pre-mRNA in eukaryotic cells. It is a core component of the spliceosome, the complex responsible for removing introns from pre-mRNA transcripts. U6 RNA is involved in the catalysis of the splicing reaction and helps in the formation of the active site of the spliceosome, facilitating the precise excision of introns and the ligation of exons.  & U6 spliceosomal RNA is a small nuclear RNA ( snRNA ) that plays a crucial role in the splicing of pre - mRNA in eukaryotic cells . It is a component of the spliceosome , the complex responsible for removing introns from pre - mRNA and joining exons together . U6 RNA is involved in the catalytic activity of the spliceosome and is essential for the recognition of splice sites and the formation of the active site for splicing . It also interacts with other snRNAs. \\
        \hline
    \end{tabular}
    \caption{Comparison of Ground Truth, GPT4o, Olmo@ft, and OmniDNA@ft in RNA type determination (Example 3)}
    \label{tab:dna2func-3}
\end{table}

\end{document}